 \documentclass[final,3p,times,twocolumn]{elsarticle}

\usepackage{amssymb}

\journal{Ad hoc Networks}

\usepackage{url}
\usepackage{balance}
\usepackage{amsfonts}
\usepackage{amsmath}
\usepackage{amsthm}
\newtheorem{definition}{Definition}
\newtheorem{lemma}{Lemma}
\newtheorem{theorem}{Theorem}
\usepackage[ruled,vlined,longend,linesnumbered]{algorithm2e}
\usepackage{caption}
\usepackage{subcaption}

\begin{document}

\begin{frontmatter}



\title{Wireless Charging for Weighted Energy Balance in Populations of Mobile Peers\tnoteref{icdcs}}
\tnotetext[icdcs]{A preliminary version of this paper appeared in \cite{2016icdcs}.}


\author[label2,label3]{Sotiris Nikoletseas} 
\author[label1]{Theofanis P. Raptis\corref{cor1}} 
\author[label2,label3]{Christoforos Raptopoulos}

\cortext[cor1]{Corresponding author at: Institute for Informatics and Telematics, National Research Council, Via G. Moruzzi, 1, 56124 Pisa, Italy, voice: +39 050 315 8282, email: \texttt{theofanis.raptis@iit.cnr.it}}

\address[label1]{Institute of Informatics and Telematics, National Research Council, Pisa, Italy}
\address[label2]{Department of Computer Engineering and Informatics, University of Patras, Patras, Greece}
\address[label3]{Computer Technology Institute and Press ``Diophantus'', Patras, Greece}

\begin{abstract}
Wireless energy transfer is an emerging technology that is used in networks of battery-powered devices in order to deliver energy and keep the network functional. Existing state-of-the-art studies have mainly focused on applying this technology on networks of relatively strong computational and communicational capabilities (wireless sensor networks, ad-hoc networks); also they assume energy transfer from special chargers to regular network nodes. Different from these works, we study how to efficiently transfer energy wirelessly in populations of battery-limited devices, towards prolonging their lifetime. In contrast to the state-of-the-art, we assume a much weaker population of distributed devices which are exchanging energy in a ``peer to peer'' manner with each other, without any special charger nodes. We address a quite general case of diverse energy levels and priorities in the network and study the problem of how the system can efficiently reach a weighted energy balance state distributively, under both loss-less and lossy power transfer assumptions. Three protocols are designed, analyzed and evaluated, achieving different performance trade-offs between energy balance quality, convergence time and energy efficiency.
\end{abstract}


\end{frontmatter}


\section{Introduction}
Next generation wirelessly networked populations are expected to consist of very large numbers of distributed portable devices carried by mobile agents that follow unpredictable and uncontrollable mobility patterns. Recently, there has been an increasing interest to combine near-field communication capabilities and wireless energy transfer in the same portable device, allowing mobile agents carrying the devices to wirelessly exchange energy. For example, the same antenna, designed to exploit its far-field properties for communication purposes, can be suitably configured for simultaneously realizing wireless energy transfer via its near-field properties. The near-field behavior of a pair of closely coupled transmitting and receiving dual-band printed monopole antennas (suitable for mobile phone applications) can make it possible to achieve both far-field performance and near-field power transfer efficiency (from 35\% to 10\%) for mobile phones located few centimeters apart \cite{massimo2015}. Further developments on the circuit design can render a device capable of achieving bi-directional, highly efficient wireless energy transfer and be used both as a transmitter and as a receiver \cite{prete2015}, \cite{schafer2015}. In this context, energy harvesting and wireless energy transfer capabilities are integrated, enabling each device to act on demand either as a wireless energy provider or as an energy harvester.

Populations of such devices have to operate under severe limitations in their computational power, data storage, the quality of communication and most crucially, their available amount of energy. For this reason, the efficient distributed co-operation of the agents towards achieving large computational and communication goals is a challenging task. An important goal in the design and efficient implementation of large networked systems is to save energy and keep the network functional for as long as possible \cite{rolim2011}, \cite{luo2011}. This can be achieved by using wireless energy transfer as an energy exchange enabling technology and applying interaction protocols among the agents which guarantee that the available energy in the network can be eventually distributed in a balanced way. 

Inspired by the Population Protocol model of \cite{angluin2004} and \cite{angluin2006}, we present a new model for configuring the wireless energy transfer process in networked systems of mobile agents. In contrast with the Population Protocol approach, our model assumes significantly stronger devices with complex wireless energy transfer hardware, not abstracted by computationally restricted finite-state automata.

\textbf{Our contribution}. The contribution of this paper is three-fold:
\begin{itemize}
\item We continue and extend the model of interactive wireless charging, presented in \cite{2016icdcs}. We present a problem statement regarding population weighted energy balance.
\item We consider the (quite different) cases of loss-less and lossy wireless energy transfer. We provide an upper bound on the time that is needed to reach weighted energy balance in the population at the loss-less case, and we experimentally investigate via simulations the complex impact of the energy levels diversity in the lossy case; also, we highlight several key elements of the charging procedure.
\item We provide and evaluate three interaction protocols which take into account different aspects of the charging procedure and achieve different performance trade-offs; one that is quite fast in achieving weighted energy balance in the loss-less case, another one that achieves weighted energy balance without wasting too much energy in the lossy case and a third one which gradually builds and maintains some knowledge of the energy levels in the network in an on-line manner.
\end{itemize}

\section{Related work}

Wireless energy transfer applications in networked environments have been lately investigated, especially in sensor and ad hoc networks. Numerous works suggest the employment of mobile wireless energy chargers in networks of sensor nodes, by combining energy transfer with data transmission and routing \cite{guo2013}, \cite{guo2014}, \cite{xie2012}, providing distributed and centralized solutions \cite{wang2014}, \cite{ndn}, \cite{angelopoulos2015} 
and collaborative charging schemes 
\cite{wu2015}, \cite{madhja2015}. Other works focus on 
multi-hop energy transfer in stationary networks \cite{liu2013}, \cite{rault2013}, as well as UAV-assisted charging of ground sensors \cite{griffin2012}, \cite{johnson2013}. Most of those wireless energy transfer applications have also been verified experimentally, using real device prototypes \cite{naderi2014}, \cite{peng2010}, \cite{raptis2015}. Although all those works provide nice solutions on the efficient charging of networks comprised of next generation devices, none of them investigates the peer-to-peer charging procedure in populations of mobile agents.

\section{The model}

We consider a population of $m$ mobile agents ${\cal M} = \{u_1, u_2, \ldots, u_m\}$, each one equipped with a \emph{battery cell}, a \emph{wireless energy transmitter} and a \emph{wireless energy receiver}. Every agent $u \in {\cal M}$ is assigned to a \emph{weight} $w_u$ that characterizes the importance or criticality of the agent. Whenever two agents meet (e.g. whenever their trajectory paths intersect), they can interact by exchanging energy between their respective battery cells, according to an \emph{interaction protocol} ${\cal P}$. We assume that agents are identical, that is they do not have IDs, they have the same hardware and run the same protocol ${\cal P}$. As a consequence, the \emph{state} of any agent $u \in {\cal M}$, at any time $t$, can be fully described by the \emph{energy} $E_t(u)$ available in its battery. 

More formally, we assume that time is discrete, and, at every time step $t \in \mathbb{N}$, a single pair of agents $u, u' \in {\cal M}$ is chosen for interaction. In the most general setting, interactions are planned by a \emph{scheduler} (that satisfies certain fairness conditions ensuring that all possible interactions will eventually occur), which can be used to abstract the movement of the agents. To allow for non-trivial efficiency analysis of our algorithmic solutions, in this paper, we consider a special case of fair scheduler, namely the \emph{probabilistic scheduler}, that, independently for every time step, selects a single interacting pair uniformly at random among all ${m \choose 2}$ pairs of agents in the population. 

Whenever a pair of agents $u, u'$ interact, they are able to exchange energy, by using their wireless energy equipment. Any transfer of energy $\varepsilon$ induces \emph{energy loss} $L(\varepsilon)$, due to the nature of wireless energy technology (e.g.~RF-to-DC conversion, materials and wiring used in the system, objects near the devices, etc.). For simplicity, we do not take into account energy loss due to movement or other activities of the agents, as this is besides the focus of the current paper (see also Section \ref{sec:problemdefinition}). In fact, we assume that most devices can be carried by individuals or other moving entities that have their own agenda, and thus devices interact when the latter happen to come in close proximity. We will assume that the energy loss function satisfies a linear law:

\begin{equation}
L(\varepsilon) = \beta \cdot \varepsilon, 
\end{equation}
where $\beta \in [0, 1)$ is a constant depending on the equipment. Therefore, if agents $u, u'$ interact at time $t$ and, according to the interaction protocol ${\cal P}$, agent $u$ should transfer energy $\varepsilon$ to $u'$, then the new energy levels of $u, u'$ at time $t$ become 

\begin{equation}
E_t(u) = E_{t-1}(u) - \varepsilon \quad \textrm{and} \quad E_t(u') = E_{t-1}(u') + \varepsilon-L(\varepsilon). \nonumber
\end{equation}
Furthermore, the energy levels of all other (i.e. non-interacting at time $t$) agents remain unchanged. Slightly abusing notation, we will write 

\begin{eqnarray}
(E_t(u), E_t(u')) & = & {\cal P}(E_{t-1}(u), E_{t-1}(u'), w_u, w_{u'}) \nonumber \\
& = & (E_{t-1}(u) - \varepsilon, E_{t-1}(u') + \varepsilon-L(\varepsilon)). \nonumber
\end{eqnarray}

\subsection{Problem definition and Metrics} \label{sec:problemdefinition}

We will say that a set of agents ${\cal M}$ is in \emph{weighted energy balance} if  $$ \frac{E_t(u)}{\sum_{x \in {\cal M}} E_x} = \frac{w_u}{\sum_{x \in {\cal M}} w_x}, \forall u \in \cal M. $$

In this paper, we study the following problem:

\begin{definition}[Population Weighted Energy Balance Problem]
Find an interaction protocol ${\cal P}$ for weighted energy balance at the minimum energy loss across agents in ${\cal M}$.
\end{definition}


In the present paper, we measure weighted energy balance by using the notion of \emph{total variation distance} from probability theory \cite{aldous2014} and stochastic processes \cite{LPW-book}.

\begin{definition}[Total variation distance]
Let $P, Q$ be two probability distributions defined on sample space ${\cal M}$. The total variation distance $\delta(P, Q)$ between $P$ and $Q$ is 

\begin{equation}
\delta(P, Q) \stackrel{def}{=} \frac{1}{2} \sum_{x \in {\cal M}} |P(x)-Q(x)|.
\end{equation}
\end{definition}
By standard results on total variation distance (see for example \cite{aldous2014}), we have the following equivalent expressions, which will be useful in our analysis. 

\begin{eqnarray}
\delta(P, Q) & = & \sum_{x \in {\cal M}: P(x) > Q(x)} (P(x)-Q(x)) \\
& = & \sum_{x \in {\cal M}: P(x) < Q(x)} (Q(x)-P(x)).
\end{eqnarray}

For any time $t \in \mathbb{N}$, we define the \emph{energy distribution} ${\cal E}_t$ on sample space ${\cal M}$ (i.e. the population of agents) given by 

\begin{equation}
{\cal E}_t(u) \stackrel{def}{=} \frac{E_t(u)}{E_t({\cal M})},
\end{equation}
for any $u \in {\cal M}$, where $E_t({\cal M}) = \sum_{x \in {\cal M}} E_t(x)$. Furthermore, we denote by ${\cal W}$ the \emph{weight distribution}, given by 

\begin{equation}
\mathcal{W} (u) = \frac{w_u}{\sum_{x \in {\cal M}} w_x}. 
\end{equation}

We will say that the population has weighted energy balance at most $\alpha$ at time $t$ if and only if $\delta({\cal E}_t, {\cal W}) \leq \alpha$. It is evident from the definition of our model that $\delta({\cal E}_t, {\cal W})$ is a random variable, depending on the specific distribution of energies in the population and the choices made by the probabilistic scheduler at time $t$. Therefore, we are rather interested in \emph{protocols that reduce the total variation distance on expectation with the smallest energy loss}. Furthermore, we measure the efficiency of a protocol ${\cal P}$ by the \emph{total expected energy loss} and the \emph{expected convergence time}\footnote{In our setting, the term \emph{time} is used to mean \emph{number of interactions}.} needed for the protocol to reach acceptable levels of weighted energy balance.

\section{Loss-less energy transfer}

In this section we present a very simple protocol for weighted energy balance in the case of loss-less energy transfer, i.e. for $\beta = 0$ (Protocol~\ref{prot:ows}). The protocol basically states that, whenever two agents $u, u'$ interact, they exchange energy so that fraction of their energy to their weight is equal. In particular, if $E_{t-1}(u), w_u$ (respectively $E_{t-1}(u'), w_{u'}$) is the energy level and weight of $u$ (respectively $u'$) at time $t-1$, then after $u, u'$ interact at time $t$, we have $\frac{E_t(u)}{w_u} = \frac{E_t(u')}{w_{u'}}$. Since there is no energy loss due to the transfer, we also have that $E_{t-1}(u)+E_{t-1}(u') = E_t(u)+E_t(u')$. Solving this linear system we then get $E_t(u) = \frac{w_u}{w_u+w_{u'}} (E_{t-1}(u)+E_{t-1}(u'))$ and $E_t(u') = \frac{w_{u'}}{w_u+w_{u'}} (E_{t-1}(u)+E_{t-1}(u'))$, as shown in the protocol definition.

\begin{algorithm}[h!]
\SetAlgorithmName{Protocol}{protocol}{List of Protocols}
\DontPrintSemicolon
\SetKwInOut{Input}{Input}\SetKwInOut{Output}{Output}
 \Input{Agents $u, u'$ with weights $w_u, w_{u'}$ and energy levels $\varepsilon_u, \varepsilon_{u'}$}
\begin{displaymath}
{\cal P}_{\text{OWS}}(\varepsilon_u, w_u, \varepsilon_{u'}, w_{u'}) = \left( \frac{w_u (\varepsilon_u + \varepsilon_{u'})}{w_u+w_{u'}}, \frac{w_{u'} (\varepsilon_u + \varepsilon_{u'})}{w_u+w_{u'}} \right).
\end{displaymath}
\caption{\texttt{Oblivious-Weighted-Share} ${\cal P}_{\text{OWS}}$}
\label{prot:ows}
\end{algorithm}

In the following Lemma, we show that, when all agents in the population use protocol ${\cal P}_{\text{OWS}}$, the total variation distance between the energy distribution and the distribution of weights ${\cal W} \stackrel{def}{=} \{{\cal W}(u)\}_{u \in {\cal M}} = \left\{ \frac{w_u}{\sum_{v \in {\cal M}} w_v} \right\}_{u \in {\cal M}}$ decreases in expectation. The proof not only leads to an upper bound on the time needed to reach weighted energy balance (see Theorem \ref{theorem:timebound}), but more importantly, \emph{highlights several key elements of the energy transfer process, which we exploit when designing interaction protocols for the case $\beta > 0$.}

\begin{lemma} \label{lemma:tvdupperlossless}
Let ${\cal M}$ be a population of agents using protocol ${\cal P}_{\text{OWS}}$. Assuming interactions are planned by the probabilistic scheduler and there is no loss from energy exchanges, we have that
\begin{align}
\mathbb{E}[\delta({\cal E}_{t}, {\cal W}) | {\cal E}_{t-1}] \leq \left( 1 - \frac{1}{{m \choose 2}} \right) \delta({\cal E}_{t-1}, {\cal W}).
\end{align}
\end{lemma}
\proof We first note that, since we are in the loss-less case, i.e. $L(\varepsilon)=0$, for any transfer of an amount $\varepsilon$ of energy, we have that $E_t({\cal M}) = E_0({\cal M})$, for any $t$ (i.e. the total energy amount remains the same). 

Define $\Delta_t \stackrel{def}{=} \delta({\cal E}_t, {\cal W}) - \delta({\cal E}_{t-1}, {\cal W})$ and assume that, at time $t$, agents $u, u'$ interact. By a simple observation, since the energy level of every other (non-interacting) agent remains the same, we have that
\begin{align} \label{eq:dtvchanges}
\Delta_t &= \frac{1}{2} \left( \left|\frac{w_u ({\cal E}_{t-1}(u)+{\cal E}_{t-1}(u'))}{w_u+w_{u'}}  - {\cal W}(u) \right| \right. \nonumber
\\& \left. + \left|\frac{w_{u'} ({\cal E}_{t-1}(u)+{\cal E}_{t-1}(u'))}{w_u+w_{u'}} - {\cal W}(u') \right| \right) \nonumber
\\& - \frac{1}{2} \left(\left|{\cal E}_{t-1}(u) - {\cal W}(u) \right| + \left|{\cal E}_{t-1}(u') - {\cal W}(u') \right| \right) 
\end{align} 

For any agent $x \in {\cal M}$ and time $t \geq 0$, set now $z_t(x) \stackrel{def}{=} {\cal E}_t(x) - {\cal W}(x)$. Let also $A^+_t \subseteq {\cal M}$ (resp. $A^-_t, A^=_t$, ) be the set of agents such that $z_t(x)$ is positive (resp. negative and equal to 0). 

Depending on relative values of the numbers $z_{t-1}(x), x \in {\cal M}$, we now distinguish the following cases:  

\begin{description}

\item[Case I:] $\, z_{t-1}(u) z_{t-1}(u') \geq 0$. In particular, if both $z_{t-1}(u), z_{t-1}(u')$ are non-negative, notice that $z_t(u) = \frac{w_u ({\cal E}_{t-1}(u)+{\cal E}_{t-1}(u'))}{w_u+w_{u'}}  - {\cal W}(u) \geq \frac{w_u}{w_u+w_{u'}} \left( {\cal W}(u) + {\cal W}(u')\right) - {\cal W}(u) \geq 0$. Similarly, $z_t(u') \geq 0$. In the same way we can prove that, if both $z_{t-1}(u), z_{t-1}(u')$ are negative, then $z_t(u) < 0$ and $z_t(u') <0$. However, notice that, either if $z_{t-1}(u), z_{t-1}(u')$ are both non-negative or if both are positive, from equation (\ref{eq:dtvchanges}) we get that $\Delta_t = 0$.  

\item[Case II: ]  $\,\,\, z_{t-1}(u) z_{t-1}(u') < 0$ and $|z_{t-1}(u)| \geq |z_{t-1}(u')|$. Assume that $z_{t-1}(u) > 0$, so $z_{t-1}(u') < 0$. By assumption we then have that $|{\cal E}_{t-1}(u) - {\cal W}(u)| \geq |{\cal E}_{t-1}(u') - {\cal W}(u')|$, or equivalently ${\cal E}_{t-1}(u) - {\cal W}(u) \geq {\cal W}(u') - {\cal E}_{t-1}(u')$, and after rearranging we get ${\cal E}_{t-1}(u) + {\cal E}_{t-1}(u') \geq {\cal W}(u) + {\cal W}(u')$. By multiplying both hand sides of the last inequality by $\frac{w_u}{w_u+w_{u'}}$ (respectively by $\frac{w_{u'}}{w_u+w_{u'}}$) we get that $z_t(u) \geq 0$ (respectively $z_t(u') \geq 0$). Therefore, from equation (\ref{eq:dtvchanges}) we get that $\Delta_t = {\cal E}_{t-1}(u') - {\cal W}(u') = z_{t-1}(u')$. 

The situation $z_{t-1}(u) < 0$ (in which case $z_{t-1}(u') > 0$) is similar, with the inequalities and signs reversed. In particular, we now have that $z_t(u) \leq 0$ and $z_t(u') \leq 0$. Therefore, from equation (\ref{eq:dtvchanges}) we get that $\Delta_t = {\cal W}(u') - {\cal E}_{t-1}(u') = -z_{t-1}(u')$.

Putting it all together, when $z_{t-1}(u) z_{t-1}(u') < 0$ and $|z_{t-1}(u)| \geq |z_{t-1}(u')|$, we have that $\Delta_t = - |z_{t-1}(u')|$.

\item[Case III:] $\,\,\,\,\, z_{t-1}(u) z_{t-1}(u') < 0$ and $|z_{t-1}(u)| < |z_{t-1}(u')|$. This case is symmetrical to Case II, with the roles of $u, u'$ exchanged. Therefore, $\Delta_t = - |z_{t-1}(u)|$.

\end{description}

Since interactions are planned by the probabilistic scheduler, i.e. any specific pair $u, u'$ of agents is chosen for interaction at time $t$ with probability $\frac{1}{{m \choose 2}}$, by linearity of expectation and equation (\ref{eq:dtvchanges}), we get


\begin{align}
\mathbb{E}[\Delta_t | {\cal E}_{t-1}] = 
-\frac{1}{{m \choose 2}} \left(\Big[\sum_{x \in {\cal M}} |z_{t-1}(x)| \cdot |\{y: |z_{t-1}(y)| \right.\nonumber\\
\left. \geq |z_{t-1}(x)|, z_{t-1}(x) z_{t-1}(y) < 0\}|\Big] -\right. \nonumber\\
- \left. \Big[\sum_{x \in {\cal M}} \frac{1}{2} |z_{t-1}(x)| \cdot |\{y: |z_{t-1}(y)| \right. \nonumber\\
= \left. |z_{t-1}(x)|, z_{t-1}(x) z_{t-1}(y) < 0\}| \vphantom{\sum_{x \in {\cal M}}} \Big]\right) \label{eq:average}
\end{align}
where we subtracted $\sum_{x \in {\cal M}} \frac{1}{2} |z_{t-1}(x)| \cdot |\{y: |z_{t-1}(y)| = |z_{t-1}(x)|, z_{t-1}(x) z_{t-1}(y) < 0\}|$ from the above sum, since the contribution $-|z_{t-1}(x)|$ of agent $x$ is counted twice for agents $x, y$ such that $|z_{t-1}(x)| = |z_{t-1}(y)|$ (once for $x$ and once for $y$). Notice also that, in the above sum, we can ignore agents $x \in A^=_{t-1}$, since their contribution to $\mathbb{E}[\Delta_t | {\cal E}_{t-1}]$ is 0. 

In order to give a formula for $\mathbb{E}[\Delta_t | {\cal E}_{t-1}]$ that is easier to handle, consider a complete ordering $\sigma_{t-1}$ of the agents $x \in A^+_{t-1} \cup A^-_{t-1}$ in increasing value of $|z_{t-1}(x)|$, breaking ties arbitrarily. We will write $x <_{\sigma_{t-1}} y$ if agent $x$ is ``to the left'' of agent $y$ in $\sigma_{t-1}$, or equivalently $\sigma_{t-1}(x) < \sigma_{t-1}(y)$. We can then see that, the contribution of an agent $x \in A^+_{t-1}$ (resp. $x \in A^-_{t-1}$) to $\mathbb{E}[\Delta_t | {\cal E}_{t-1}]$ is $-|z_{t-1}(x)|$ multiplied by the number of agents in $A^-_{t-1}$ (resp. $A^+_{t-1}$) that are ``to the right'' of $x$ in $\sigma_{t-1}$ (i.e. agents $y$ such that $x <_{\sigma_{t-1}} y$, for which $z_{t-1}(x) z_{t-1}(y) < 0$). Therefore, equation (\ref{eq:average}) becomes
\begin{align}
&\mathbb{E}[\Delta_t | {\cal E}_{t-1}] = \\\nonumber &-\frac{1}{{m \choose 2}} \sum_{x \in {\cal M}} |z_{t-1}(x)| \cdot |\{y: x<_{\sigma_{t-1}} y, z_{t-1}(x) z_{t-1}(y) < 0\}|.
\end{align}
Assume now, without loss of generality, that the ``rightmost'' agent in $\sigma_{t-1}$ is some $y^* \in A^+_{t-1}$. By the above equation, we then have that the contribution $-|z_{t-1}(x)|$ of every agent $x \in A^-_{t-1}$ is counted at least once (because of $y^*$, since $x<_{\sigma_{t-1}} y^*$ and $z_{t-1}(x) z_{t-1}(y^*) < 0$). Therefore,
\begin{equation}
\mathbb{E}[\Delta_t | {\cal E}_{t-1}] \leq -\frac{1}{{m \choose 2}} \sum_{x \in A^-_{t-1}} |z_{t-1}(x)|.
\end{equation}
But using a standard result on total variation distance (see for example \cite{aldous2014}), we have that 

\begin{equation}
\delta({\cal E}_{t-1}, {\cal W}) = \sum_{x \in A^-_{t-1}} |z_{t-1}(x)| = \sum_{y \in A^+_{t-1}} |z_{t-1}(y)|
\end{equation}
Therefore, we have that $\mathbb{E}[\Delta_t | {\cal E}_{t-1}] \leq -\frac{1}{{m \choose 2}} \delta({\cal E}_{t-1}, {\cal W})$, which completes the proof.\qed

It is worth noting that the upper bound of Lemma \ref{lemma:tvdupperlossless} is tight when the distribution of energies is such that there is only one agent $u$ with energy above or below his weighted average ${\cal W}(u) = \frac{w_u}{\sum_{v \in {\cal M}} w_v}$.

We now use Lemma \ref{lemma:tvdupperlossless} to prove that protocol ${\cal P}_{\text{OWS}}$ is quite fast in achieving weighted energy balance in the loss-less case.

\begin{theorem} \label{theorem:timebound}
Let ${\cal M}$ be a population of agents using protocol ${\cal P}_{\text{OWS}}$. Let also $\tau_0(c)$ be the time after which $\mathbb{E}[\delta({\cal E}_{\tau_0(c)}, {\cal W})] \leq c$, assuming interactions are planned by the probabilistic scheduler and there is no loss from energy exchanges. Then $\tau_0(c) \leq {m \choose 2} \ln\left(\frac{\delta({\cal E}_0, {\cal W})}{c} \right)$, where $\delta({\cal E}_0, {\cal W})$ is the total variation distance between the initial energy distribution and the distribution of weights ${\cal W} \stackrel{def}{=} \{{\cal W}(u)\}_{u \in {\cal M}} = \left\{ \frac{w_u}{\sum_{v \in {\cal M}} w_v} \right\}_{u \in {\cal M}}$.
\end{theorem}
\proof Taking expectations in the upper bound inequality from Lemma \ref{lemma:tvdupperlossless}, we have that $\mathbb{E}[\mathbb{E}[\delta({\cal E}_{t}, {\cal W}) | {\cal E}_{t-1}]] \leq \left( 1 - \frac{1}{{m \choose 2}} \right) \mathbb{E}[\delta({\cal E}_{t-1}, {\cal W})]$, or equivalently

\begin{equation}
\mathbb{E}[\delta({\cal E}_t, {\cal W})] \leq \left( 1 - \frac{1}{{m \choose 2}} \right) \mathbb{E}[\delta({\cal E}_{t-1}, {\cal W})].
\end{equation}
Iterating the above inequality, we then have that
\begin{equation}
\mathbb{E}[\delta({\cal E}_t, {\cal W})] \leq \left(1 - \frac{1}{{m \choose 2}} \right)^t \delta({\cal E}_0, {\cal W}) \leq e^{-\frac{t}{{m \choose 2}}} \delta({\cal E}_0, {\cal W}).
\end{equation}
Consequently, for any $t \geq {m \choose 2} \ln\left(\frac{\delta({\cal E}_0, {\cal W})}{c} \right)$, we have that $\mathbb{E}[\delta({\cal E}_t, {\cal W})] \leq c$, which concludes the proof. \qed

\section{Energy transfer with loss} \label{sec:lossy}

In this section we consider the more natural case where every transfer of energy $\varepsilon$ induces energy loss $L(\varepsilon) = \beta \varepsilon$, for some $0 <\beta < 1$. The main technical difficulty that arises in this case when considering the total variation distance change $\Delta(t) = \delta({\cal E}_t, {\cal W}) - \delta({\cal E}_{t-1}, {\cal W})$ is that any energy transfer between agents $u$ and $u'$ affects also the relative energy levels of non-interacting agents $x$ in comparison with their respective target values $\frac{w_x}{\sum_{y \in {\cal M}} w_y}$. More precisely, after $u, u'$ exchange energy $\varepsilon$ at time $t$, we have $E_t({\cal M}) = E_{t-1}({\cal M}) - \beta \varepsilon$. Therefore, for any non-interacting agent $x$ at time $t$, we have $|z_t(x)| \stackrel{def}{=} \left|{\cal E}_t(x) - \mathcal{W}(u) \right| = \left|\frac{E_t(x)}{E_t({\cal M})} - \mathcal{W}(u) \right| \neq \left|\frac{E_{t-1}(x)}{E_{t-1}({\cal M})} - \mathcal{W}(u) \right| \stackrel{def}{=} |z_{t-1}(x)|$. As a consequence, straightforward generalizations of simple protocols like ${\cal P}_{\text{OWS}}$ do not perform up to par in this case. 

In particular, there are specific worst-case distributions of energies for which the total variation distance increases on expectation after any significant energy exchange. As a fictitious example, consider a population of $m$ agents, for some $m > \frac{2+\beta}{\beta}$ and $w_u = w_{u'}, \forall u, u' \in \cal M$. Furthermore, suppose that agents $u_i$, for $i \in [m-1]$ have energy $\varepsilon_{u_i} = m$, while agent $u_m$ has energy $\varepsilon_{u_m} = 0$ at his disposal. The total variation distance in this example is $\frac{1}{m}$. Without loss of generality, consider now a variation of ${\cal P}_{\text{OWS}}$ (adapted from the original version for $\beta=0$ to the case of $\beta>0$) according to which, whenever two agents $u, u'$ interact, the agent with the largest amount of energy transfers $\frac{|\varepsilon_u - \varepsilon_{u'}|}{2}$ energy to the other.\footnote{Nevertheless, it is not hard to see that other variations of ${\cal P}_{\text{OWS}}$ have similar problems.} We now have that, after any significant energy exchange step, i.e. an interaction of $u_m$ with any other agent, say $x$, according to protocol ${\cal P}_{\text{OWS}}$, the new energy level of $u_m$ becomes $\frac{m}{2}(1-\beta)$, the new energy level of $x$ becomes $\frac{m}{2}$, while the energy levels of all other agents remain the same. Therefore, the new total energy in the population becomes $m^2-m-\frac{\beta}{2} m$, and the new total variation distance becomes\footnote{The total variation distance consists of 2 terms, since there are only 2 agents with energy levels below the average.} $\left( \frac{1}{m} - \frac{1}{2m-2-\beta}\right) + \left( \frac{1}{m} - \frac{1}{2m-2-\beta} (1-\beta)\right) = \frac{2}{m} - \frac{2-\beta}{2m-2-\beta}$, which is strictly larger than $\frac{1}{m}$, for any $m > \frac{2+\beta}{\beta}$. We conclude that, in this example, the total variation distance increases also in expectation, as any interaction between pairs of agents that do not contain $u_m$ does not change the energy distribution. 

\begin{figure}[t!]
\centering
\includegraphics[width=0.30\textwidth]{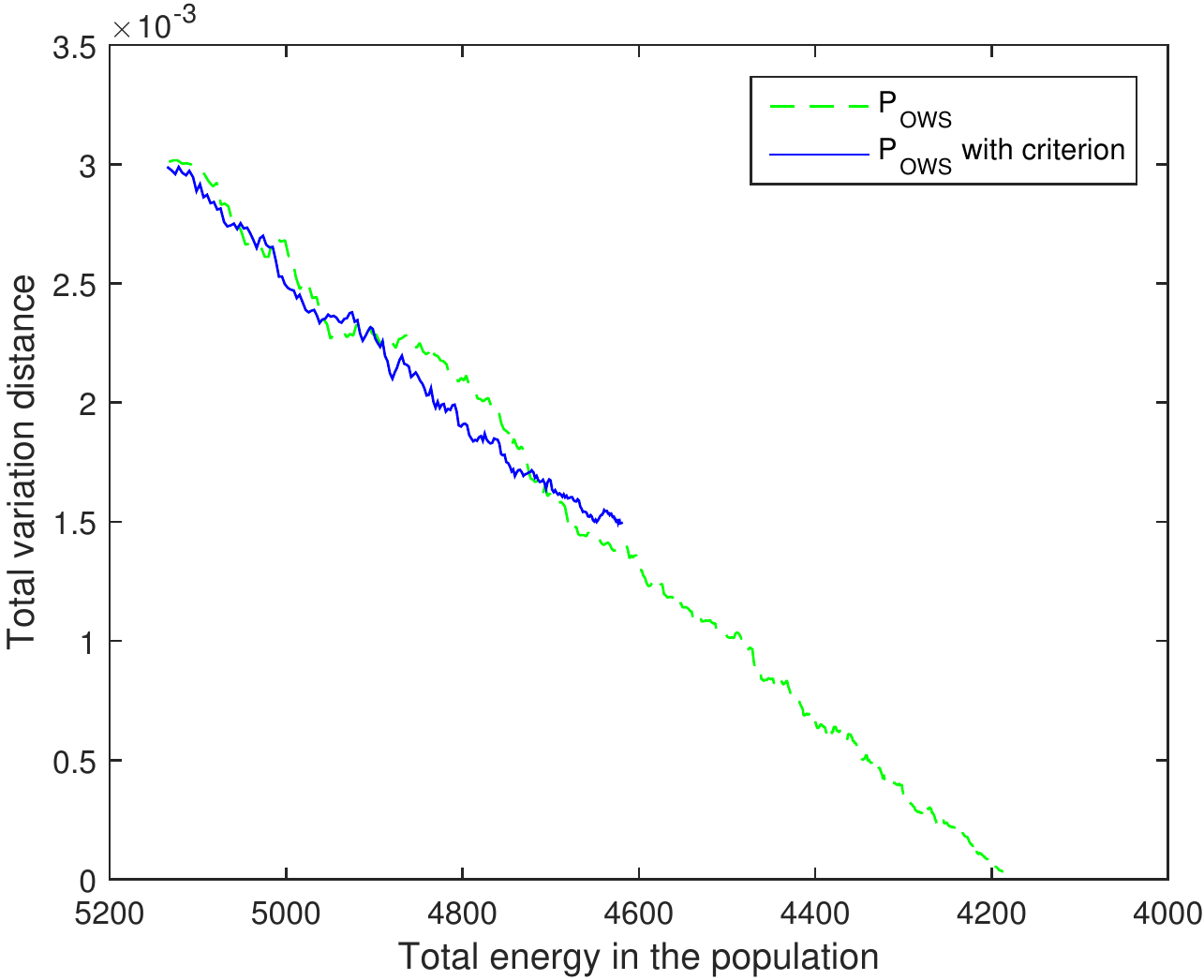}
\caption{Efficiency of the ${\cal P}_{\text{OWS}}$ protocol applying the criterion that allows transfers between agents with significantly different energy levels, compared to the standard protocol version.}
        \label{fig:attempt}
\end{figure}

It is worth noting that, even though the above example is fictitious, our experiments verify our intuition that ${\cal P}_{\text{OWS}}$ is not very suitable for weighted energy balance in the case of lossy energy transfer. In particular, it seems that the energy lost with every step does not contribute sufficiently to the reduction of total variation distance between the distribution of energies and the uniform distribution. Our first attempt to overcome this problem was to only allow energy transfers between agents whose energy levels differ significantly. However, this did not solve the problem either (see our experimental results in Fig.~\ref{fig:attempt}), mainly because of interactions between agents that are both below the average energy. As our main contribution, we present in Subsections \ref{sec:PST} and \ref{sec:POA} two interaction protocols that seem to make the most of the energy lost in every step. 

\subsection{The protocol \texttt{Small-Weighted-Transfer} ${\cal P}_{\text{SWT}}$} \label{sec:PST}

The protocol \texttt{Small-Weighted-Transfer} ${\cal P}_{\text{SWT}}$ (Prot.~\ref{prot:st}) suggests having only small energy transfers between interacting agents that are proportional to the difference of their relative energies $\frac{\varepsilon_u}{w_u}$. In particular we assume that $d\varepsilon$ is a universal constant known to all agents and corresponds to a very small amount of energy. Whenever agents $u$ and $u'$ interact and the relative energy level of $u$ is higher than the relative energy level of $u'$, then $u$ transfers energy $\left| \frac{\varepsilon_u}{w_u} - \frac{\varepsilon_{u'}}{w_{u'}} \right| d\varepsilon$ to $u'$. The main idea behind ${\cal P}_{\text{SWT}}$ is that pathological energy distributions that lead to increment of the total variation distance $\delta({\cal E}, {\cal W})$ occur rarely on average. Additionally, by allowing only small energy transfers, we avoid ``overshooting'' cases where devices below (respectively above) their target energy value are overcharged (respectively transfer more energy than they should). Even though this idea can be wasteful on time, we provide experimental evidence that it achieves weighted energy balance without wasting too much energy.

\begin{algorithm}[t!]
\SetAlgorithmName{Protocol}{protocol}{List of Protocols}
\DontPrintSemicolon
\SetKwInOut{Input}{Input}\SetKwInOut{Output}{Output}
 \Input{Agents $u, u'$ with weights $w_u, w_{u'}$ and energy levels $\varepsilon_u, \varepsilon_{u'}$}
 Set $$\phi = \left| \frac{\varepsilon_u}{w_u} - \frac{\varepsilon_{u'}}{w_{u'}} \right|$$\;
\If{$\frac{\varepsilon_u - \phi d \varepsilon}{w_u} \geq \frac{\varepsilon_{u'} + \phi d \varepsilon}{w_{u'}}$}{${\cal P}_{\text{SWT}}(\varepsilon_u, w_u, \varepsilon_{u'}, w_{u'}) = \left( \varepsilon_u - \phi d\varepsilon, \varepsilon_{u'} + (1-\beta) \phi d\varepsilon \right)$}
\ElseIf{$\frac{\varepsilon_u + \phi d \varepsilon}{w_u} < \frac{\varepsilon_{u'} - \phi d \varepsilon}{w_{u'}}$}{${\cal P}_{\text{SWT}}(\varepsilon_u, w_u, \varepsilon_{u'}, w_{u'}) = \left( \varepsilon_u + (1-\beta) \phi d\varepsilon, \varepsilon_{u'} - \phi d\varepsilon \right)$}
\Else{do nothing.}
%
 \caption{\texttt{Small-Weighted-Transfer} ${\cal P}_{\text{SWT}}$}
\label{prot:st}
\end{algorithm}

We prove the following lemma concerning the total variation distance change in a population of agents that use protocol ${\cal P}_{\text{SWT}}$. Due to the difficulty of the analysis, we select a specific assignment of weights to the agents by setting $w_u = 1, \forall u \in \cal M$. Consequently, the weight distribution $\cal W$ becomes equal to the uniform distribution, denoted as $\mathcal{U}$.

\begin{lemma} \label{lemma:tvdupperlossy}
Let ${\cal M}$ be a population of chargers using protocol ${\cal P}_{\text{SWT}}$. Given any distribution of energy ${\cal E}_{t-1}$, let $|A^+_{t-1}|$ (respectively $|A^-_{t-1}|$) be the number of agents with available energy above (respectively below) the current average. Assuming interactions are planned by the probabilistic scheduler, we have that
\begin{equation}
\mathbb{E}[\Delta_t | {\cal E}_{t-1}] \leq \frac{4}{E_t({\cal M})} \left( \beta - \frac{|A^+_{t-1}| \cdot |A^-_{t-1}|}{m(m-1)}\right).
\end{equation}
\end{lemma}
\proof We will use the notation from Lemma \ref{lemma:tvdupperlossless}. Furthermore, let $a^+ = |A^+_{t-1}|, a^- = |A^-_{t-1}|$ and $a^= = |A^=_{t-1}|$. Assume without loss of generality, that at time $t$, the probabilistic scheduler selects agents $u, u'$, such that $E_{t-1}(u) > E_{t-1}(u')-d\varepsilon$. Therefore, according to ${\cal P}_{\text{SWT}}$, agent $u$ transfers energy $d\varepsilon$ to $u'$, and so $E_t(u) = E_{t-1}(u) - d\varepsilon$ and $E_{t}(u') = E_{t-1}(u') + (1-\beta) d\varepsilon$. The energy level of every other charger remains unchanged. Furthermore, the new total energy in the population is
\begin{equation}
E_t({\cal M}) = E_{t-1}({\cal M}) - \beta d\varepsilon.
\end{equation}
A crucial observation for the analysis is that, since ${\cal P}_{\text{SWT}}$ only allows transfers of very small amounts of energy, after any useful interaction (i.e. interactions that change the distribution of energy in the population), the only agents that can potentially change the relative position of their energy levels to the average energy are those in $A^=_{t-1}$. 

We now distinguish the following cases:

\begin{description}

\item[Case I:]  \hspace{0.2cm} For any $x \in (A^+_{t-1} \cup A^=_{t-1}) \backslash \{u, u'\}$, we have that $z_{t-1}(x) \geq 0$, so $z_{t}(x) = \frac{E_{t-1}(x)}{E_{t-1}({\cal M}) - \beta d\varepsilon} - \frac{1}{m} > 0$. Therefore,
\begin{align}
|z_t(x)| =  \frac{E_{t-1}(x)}{E_{t-1}({\cal M}) - \beta d\varepsilon} - \frac{1}{m} \\
 =  \frac{E_{t-1}({\cal M})}{E_t({\cal M})} |z_{t-1}(x)| + \beta \frac{1}{m} \frac{1}{E_t({\cal M})} d\varepsilon.
\end{align}
\item[Case II:] \hspace{0.2cm} For any $x \in A^-_{t-1} \backslash \{u, u'\}$, we have that $z_{t-1}(x) < 0$, so $z_{t}(x) = \frac{E_{t-1}(x)}{E_{t-1}({\cal M}) - \beta d\varepsilon} - \frac{1}{m} < 0$, for any very small energy transfer $d\varepsilon$. Therefore,
\begin{align}
|z_t(x)|  =  \frac{E_{t-1}(x)}{E_{t-1}({\cal M}) - \beta d\varepsilon} - \frac{1}{m} \\
 =  \frac{E_{t-1}({\cal M})}{E_t({\cal M})} |z_{t-1}(x)| - \beta \frac{1}{m} \frac{1}{E_t({\cal M})} d\varepsilon.
\end{align}
\item[Case III:] \hspace{0.2cm} If $u \in A^-_{t-1} \cup A^=_{t-1}$, then $z_{t-1}(u) \leq 0$, so $z_{t}(u) = \frac{E_{t-1}(x) - d\varepsilon}{E_{t-1}({\cal M}) - \beta d\varepsilon} - \frac{1}{m} < 0$, since $E_{t-1}(u) \leq E_{t-1}({\cal M})$ and $\beta \in (0, 1)$. Furthermore, by assumption, $z_{t-1}(u') < 0$, and also (by the conditions of ${\cal P}_{\text{SWT}}$), $z_{t}(u') = \frac{E_{t-1}(u') + (1-\beta) d \varepsilon}{E_{t-1}({\cal M}) -\beta d\varepsilon} - \frac{1}{m} \leq \frac{E_{t-1}(x) - d\varepsilon}{E_{t-1}({\cal M}) - \beta d\varepsilon} - \frac{1}{m} < 0$. Therefore,
\begin{align}
|z_t(u)| + |z_t(u')| = \nonumber \\
=  \frac{1}{m} - \frac{E_{t-1}(u) - d\varepsilon}{E_{t-1}({\cal M}) - \beta d\varepsilon} + \frac{1}{m} \nonumber \\ 
- \frac{E_{t-1}(u') + (1-\beta)d \varepsilon}{E_{t-1}({\cal M}) - \beta d\varepsilon} \\
 =  \frac{E_{t-1}({\cal M})}{E_t({\cal M})} (|z_{t-1}(u)|+|z_{t-1}(u')|) \nonumber \\
 + \beta \left(1 - \frac{2}{m} \right) \frac{1}{E_t({\cal M})} d\varepsilon.
\end{align}
\item[Case IV:] \hspace{0.2cm} If $u' \in A^+_{t-1} \cup A^=_{t-1}$, then $z_{t-1}(u') \geq 0$, so $z_{t}(u') = \frac{E_{t-1}(u') + (1-\beta) d \varepsilon}{E_{t-1}({\cal M}) -\beta d\varepsilon} - \frac{1}{m} \geq 0$. Furthermore, by assumption, $z_{t-1}(u) > 0$, and also (by the conditions of ${\cal P}_{\text{SWT}}$), $z_{t}(u) = \frac{E_{t-1}(x) - d\varepsilon}{E_{t-1}({\cal M}) - \beta d\varepsilon} - \frac{1}{m} > \frac{E_{t-1}(u') + (1-\beta) d \varepsilon}{E_{t-1}({\cal M}) -\beta d\varepsilon} - \frac{1}{m} \geq 0$. Therefore,
\begin{align}
|z_t(u)| + |z_t(u')|  =  \frac{E_{t-1}(u) - d\varepsilon}{E_{t-1}({\cal M}) - \beta d\varepsilon} - \frac{1}{m} \nonumber \\
+ \frac{E_{t-1}(u') + (1-\beta)d \varepsilon}{E_{t-1}({\cal M}) - \beta d\varepsilon} - \frac{1}{m}\\
 =  \frac{E_{t-1}({\cal M})}{E_t({\cal M})} (|z_{t-1}(u)|+|z_{t-1}(u')|) \nonumber \\
 - \beta \left(1 - \frac{2}{m} \right) \frac{1}{E_t({\cal M})} d\varepsilon.
\end{align}
\item[Case V:] \hspace{0.2cm} If $u \in A^+_{t-1}$ and $u' \in A^-_{t-1}$, then, similarly to the other cases we have $|z_{t-1}(u)|>0, |z_t(u)|>0, |z_{t-1}(u')|<0$ and $|z_t(u')|<0$. Therefore,
\begin{align}
|z_t(u)| + |z_t(u')|  =  \frac{E_{t-1}(u) - d\varepsilon}{E_{t-1}({\cal M}) - \beta d\varepsilon} - \frac{1}{m} + \frac{1}{m} \nonumber \\
 - \frac{E_{t-1}(u') + (1-\beta)d \varepsilon}{E_{t-1}({\cal M}) - \beta d\varepsilon}\\
 =  \frac{E_{t-1}({\cal M})}{E_t({\cal M})} (|z_{t-1}(u)|+|z_{t-1}(u')|) \nonumber \\
 - (2-\beta) \frac{1}{E_t({\cal M})} d\varepsilon.
\end{align}
\item[Case VI:] \hspace{0.2cm} If $u, u' \in A^=_{t-1}$ there is no change in the energy distribution.

\end{description}

Furthermore, the probability that the agents $u, u'$, that are chosen for interaction by the probabilistic scheduler, are such that the conditions of case III (respectively case IV and case V) are satisfied, is $p_{\text{III}} = \frac{a^- (a^- - 1) + 2a^- a^=}{m (m-1)}$ (respectively $p_{\text{IV}} = \frac{a^+ (a^+ - 1) + 2a^+ a^=}{m (m-1)}$ and $p_{\text{V}} = \frac{2 a^- a^+}{m (m-1)}$). Putting it all together, by linearity of expectation, we have 
\begin{align}	
\mathbb{E}[\delta({\cal E}_t, {\cal U}) | {\cal E}_{t-1}]  =   \nonumber \\
 p_{\text{III}} \cdot \left( \frac{E_{t-1}({\cal M})}{E_t({\cal M})} \delta({\cal E}_{t-1}, {\cal U}) \right. \nonumber \\ 
 \left. - \frac{1}{E_t({\cal M})} \left( -\beta - \frac{\beta (a^+ + a^= - a^-)}{m}\right) d\varepsilon \right) \nonumber \\
 \quad + p_{\text{IV}} \cdot \left( \frac{E_{t-1}({\cal M})}{E_t({\cal M})} \delta({\cal E}_{t-1}, {\cal U}) \right. \nonumber \\ 
 \left. - \frac{1}{E_t({\cal M})} \left( \beta - \frac{\beta (a^+ + a^= - a^-)}{m}\right) d\varepsilon \right)  \nonumber \\
 \quad + p_{\text{V}} \cdot \left( \frac{E_{t-1}({\cal M})}{E_t({\cal M})} \delta({\cal E}_{t-1}, {\cal U}) \right. \nonumber \\
 \left. - \frac{1}{E_t({\cal M})} \left( 2-\beta - \frac{\beta (a^+ + a^= - a^-)}{m}\right) d\varepsilon \right) \nonumber \\
 =  (p_{\text{III}} + p_{\text{IV}} + p_{\text{V}}) \frac{E_{t-1}({\cal M})}{E_t({\cal M})} \delta({\cal E}_{t-1}, {\cal U})  \nonumber \\
 \quad + (p_{\text{III}} + p_{\text{IV}} + p_{\text{V}}) \frac{1}{E_t({\cal M})} \frac{\beta (a^+ + a^= - a^-)}{m} d\varepsilon  \nonumber \\
 \quad + \frac{\beta (p_{\text{III}} + p_{\text{V}}- p_{\text{IV}}) - 2 p_{\text{V}}}{E_t({\cal M})} d\varepsilon.
\end{align}
Rearranging, we have
\begin{align}
\mathbb{E}[\Delta_t | {\cal E}_{t-1}]  =  \nonumber \\
 \left((p_{\text{III}} + p_{\text{IV}} + p_{\text{V}}) \left( 1+\frac{\beta d \varepsilon}{E_t({\cal M})} \right) -1 \right)\delta({\cal E}_{t-1}, {\cal U}) + \nonumber \\
 \quad + (p_{\text{III}} + p_{\text{IV}} + p_{\text{V}}) \frac{1}{E_t({\cal M})} \frac{\beta (a^+ + a^= - a^-)}{m} d\varepsilon + \nonumber \\
 \quad + \frac{\beta (p_{\text{III}} + p_{\text{V}}- p_{\text{IV}}) - 2 p_{\text{V}}}{E_t({\cal M})} d\varepsilon
 \label{ineq:goodone}
\end{align}

By now using the fact that $p_{\text{III}}, p_{\text{IV}}, p_{\text{V}} \in [0, 1]$, $p_{\text{III}}+ p_{\text{IV}}+ p_{\text{V}} \leq 1$ and the fact that the total variation distance between any two distributions is at most 1 (see e.g. \cite{aldous2014}), we get $\mathbb{E}[\Delta_t | {\cal E}_{t-1}] \leq \frac{4}{E_t({\cal M})} \left( \beta - \frac{a^+ a^-}{m(m-1)}\right)$, which completes the proof of the Lemma. \qed

It is worth noting that the upper bound on the total variation distance change from the above Lemma is quite crude (and can be positive if $\beta$ is not small enough). However, this is mainly a consequence of our analysis; in typical situations, the upper bound that we get from inequality (\ref{ineq:goodone}) can be much smaller. For example, if the energy distribution ${\cal E}_{t-1}$ at $t-1$ is such that $|A^+_{t-1}| \approx |A^-_{t-1}| \approx \frac{m}{2}$, (\ref{ineq:goodone}) gives the bound $\mathbb{E}[\Delta_t | {\cal E}_{t-1}] \leq -\frac{1-\beta}{E_t({\cal M})} d\varepsilon$, which is negative for any $\beta \in (0, 1)$. This is also verified by our experimental evaluation of ${\cal P}_{\text{SWT}}$. Nevertheless, the upper bound that we get from Lemma \ref{lemma:tvdupperlossy} highlights key characteristics of the interactive energy transfer process as we pass from loss-less (i.e. $\beta=0$) to loss-y energy transfer (i.e. $\beta>0$).

\subsection{The protocol \texttt{Online-Weighted-Average} ${\cal P}_{\text{OWA}}$} \label{sec:POA}

By the analysis of the expected total variation distance change in Lemma \ref{lemma:tvdupperlossless} for energy transfer without losses, we can see that the total variation distance decreases when the interacting agents have relative energy levels that are on different sides of their respective target values. Using the notation form the proof of Lemma \ref{lemma:tvdupperlossless}, if agents $u, u'$ interact at time $t$, then we must either have $u \in A^+_{t-1}$ and $u' \in A^-_{t-1}$, or $u \in A^-_{t-1}$ and $u' \in A^+_{t-1}$, in order for the total variation distance $\delta({\cal E}_t, {\cal W})$ to drop below $\delta({\cal E}_{t-1}, {\cal W})$. The situation becomes more complicated when there are losses in energy transfers.

In view of the above, an ideal interaction protocol would only allow energy transfers between agents $u, u'$ for which $z_{t-1}(u)z_{t-1}(u') < 0$. In particular, this implies that, at any time $t$, each agent $x$ would need to know the sign of $z_t(x) = \frac{E_t(x)}{E_t({\cal M})} - \frac{w_u}{\sum_{x \in \cal{M}}w_x}$, which is possible if $x$ knows (in addition to its own energy level and weight) its target weighted average energy $\frac{w_u}{\sum_{x \in \cal{M}}w_x} E(\cal{M})$ in the population. However, this kind of global knowledge is too powerful in our distributed model, since we assume that agents are independent and identical with each other. In particular, in our model, not only are agents not aware of other agents they have not yet interacted with, but also, that they have no way of knowing whether they have met with another agent at some point in the past. 

\begin{algorithm}[h!]
\SetAlgorithmName{Protocol}{protocol}{List of Protocols}
\DontPrintSemicolon
\SetKwInOut{Input}{Input}\SetKwInOut{Output}{Output}
 \Input{Agents $u, u'$ with weights $w_u, w_{u'}$ and energy levels $\varepsilon_u, \varepsilon_{u'}$}
Set $\text{nrg}(u) = \text{nrg}(u) + \varepsilon_{u'}$ and $\text{nrg}(u') = \text{nrg}(u') + \varepsilon_{u}$.\;
Set $\text{wt}(u) = \text{wt}(u) + w_{u'}$ and  $\text{wt}(u') = \text{wt}(u') + w_{u}$.\;
\If{($\varepsilon_u >  \frac{w_u}{\textnormal{wt}(u)}\textnormal{nrg}(u)$ AND $\varepsilon_u' \leq  \frac{w_{u'}}{\textnormal{wt}(u')}\textnormal{nrg}(u')$) OR ($\varepsilon_u \leq  \frac{w_u}{\textnormal{wt}(u)}\textnormal{nrg}(u)$ AND $\varepsilon_u' >  \frac{w_{u'}}{\textnormal{wt}(u')}\textnormal{nrg}(u')$)}{
Set $\delta = \left|\frac{w_{u'}\varepsilon_{u} - w_{u}\varepsilon_{u'}}{w_{u} + w_{u'}} \right|$\;
{\If{$\frac{\varepsilon_u}{w_u} > \frac{\varepsilon_{u'}}{w_{u'}}$}{
${\cal P}_{\text{OWA}}(\varepsilon_u, w_u, \varepsilon_{u'}, w_{u'}) = \left( \varepsilon_u - \delta, \varepsilon_{u'} + (1-\beta)\delta \right)$
}
\ElseIf{$\frac{\varepsilon_u}{w_u} \leq \frac{\varepsilon_{u'}}{w_{u'}}$}{
${\cal P}_{\text{OWA}}(\varepsilon_u, w_u, \varepsilon_{u'}, w_{u'}) = \left( \varepsilon_u + (1-\beta)\delta, \varepsilon_{u'} - \delta \right)$
}
}
}
\Else{do nothing.}
 \caption{\texttt{Online-Weighted-Average} ${\cal P}_{\text{OWA}}$}
\label{prot:oa}
\end{algorithm}

The main idea behind our interaction protocol ${\cal P}_{\text{OWA}}$ (Prot.~\ref{prot:oa}) is that, even in our weak model of local interactions, agents are still able to compute local estimates of their respective target energy levels based on the energy levels and weights of agents they interact with. To do this, every agent needs to keep track of the cumulative energy and cumulative weights of agents she comes across\footnote{Notice that, by definition, agents cannot store all weights and all energy levels of other agents separately. Indeed, this contradicts our assumption that agents are identical.}. This is accomplished by having each agent $x \in {\cal M}$ maintaining two local registers: (a) $\text{wt}(x)$ which is used to store the cumulative weight (note that $\text{wt}(x)$ may include more than once the weight of a particular agent, since agents are assumed identical), and (b) $\text{nrg}(x)$, which is used to store the cumulative energy. For each agent $x \in {\cal M}$, $\text{wt}(x)$ is initialized to $w_x$, and $\text{nrg}(x)$ is initialized to $E_0(x)$. 
The exact amount $\delta$ of energy transfer from agent $u$ to agent $u'$, is calculated from the equation
\begin{equation}
\frac{\varepsilon - \delta}{w_u} = \frac{\varepsilon_{u'} + \delta}{w_{u'}}
\end{equation}
which corresponds to the intended energy exchange when there is no loss.


It is worth noting that ${\cal P}_{\text{OWA}}$ may not perform as expected in the general case where interactions are planned by a potentially adversarial scheduler, because the local estimates kept by agents for the average can be highly biased. On the other hand, in our experimental evaluation, we show that ${\cal P}_{\text{OWA}}$ outperforms both ${\cal P}_{\text{OWS}}$ and ${\cal P}_{\text{SWT}}$ when agent interactions are planned by the probabilistic scheduler. Furthermore, it is much faster than ${\cal P}_{\text{SWT}}$ in terms of the expected number of useful interactions (i.e. interactions that change the energy distribution in the population) needed to reach weighted energy balance.

\section{Performance evaluation}

\begin{figure}[t!]
\centering
\includegraphics[width=0.30\textwidth]{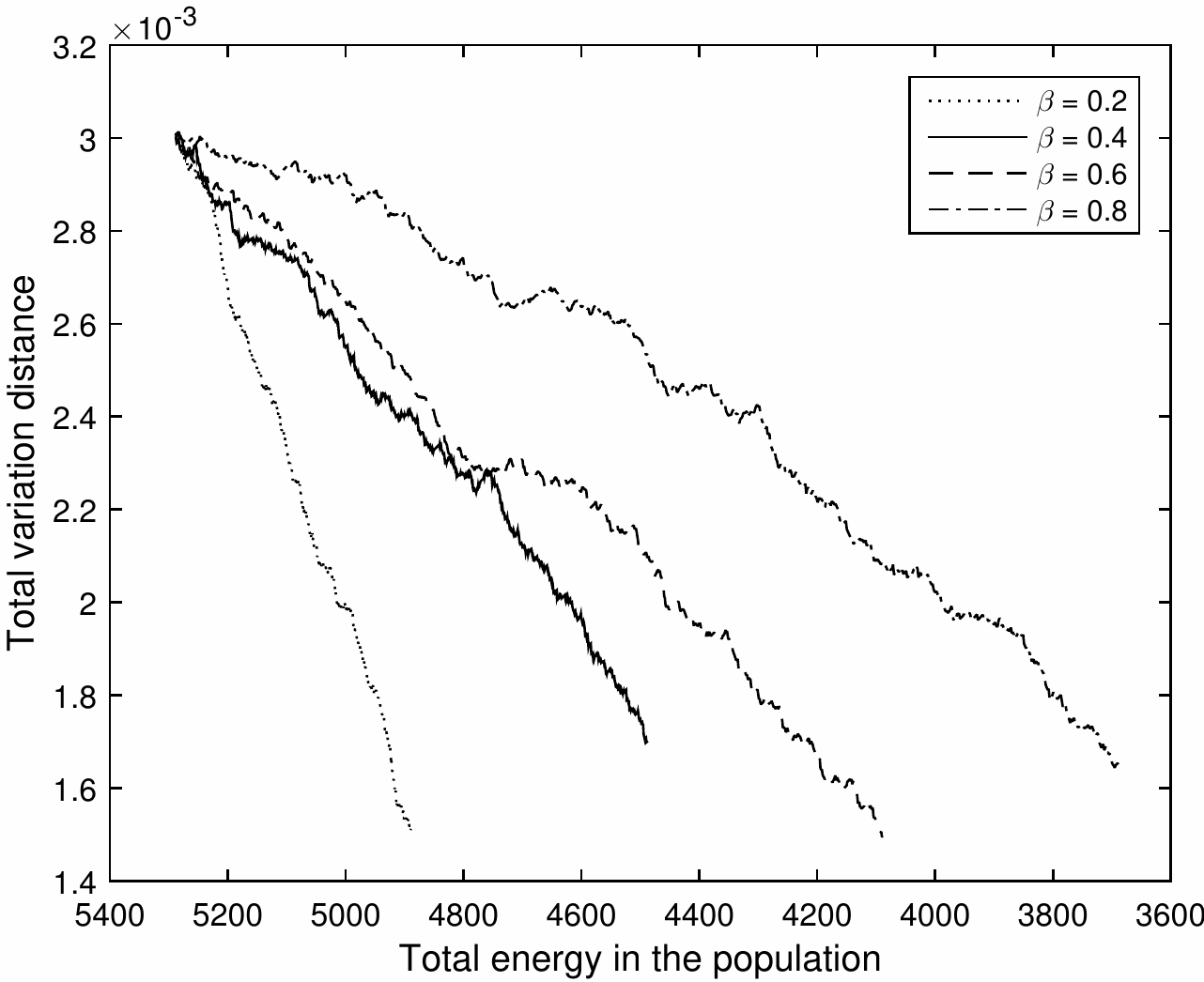}
\caption{Performance of ${\cal P}_{\text{SWT}}$ for different values of $\beta$. Different loss functions affect the performance of the protocol.}
\label{fig:beta}
\end{figure}

\begin{figure*}[t!]
        \centering
        \begin{subfigure}[b]{0.33\textwidth}
                \includegraphics[width=\textwidth]{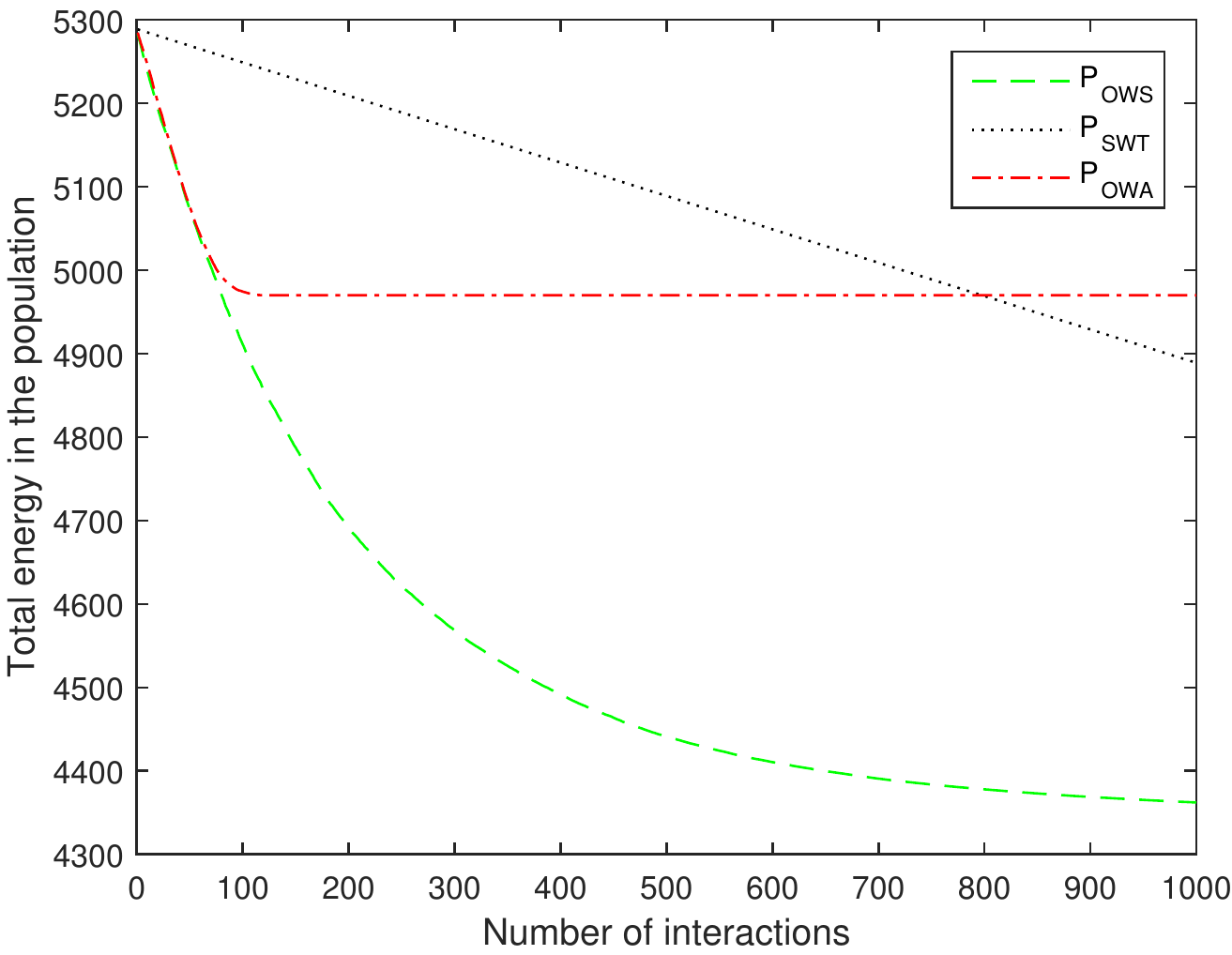}
                \caption{Energy loss, $\beta = 0.2$.}
                \label{fig:loss1}
        \end{subfigure}%
        \begin{subfigure}[b]{0.33\textwidth}
                \includegraphics[width=\textwidth]{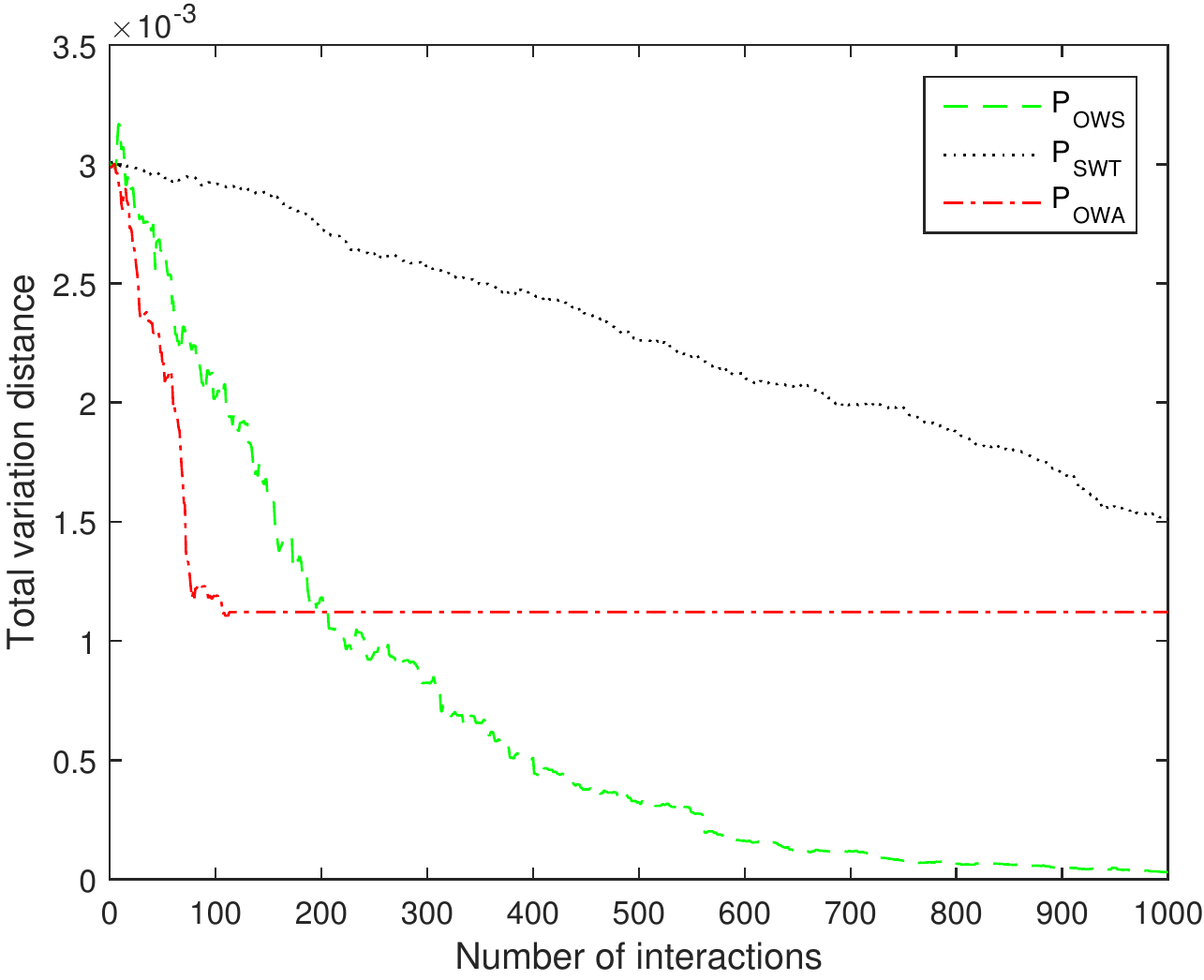}
                \caption{Weighted energy balance, $\beta = 0.2$.}
                \label{fig:balance1}
        \end{subfigure}%
        \begin{subfigure}[b]{0.33\textwidth}
                \includegraphics[width=\textwidth]{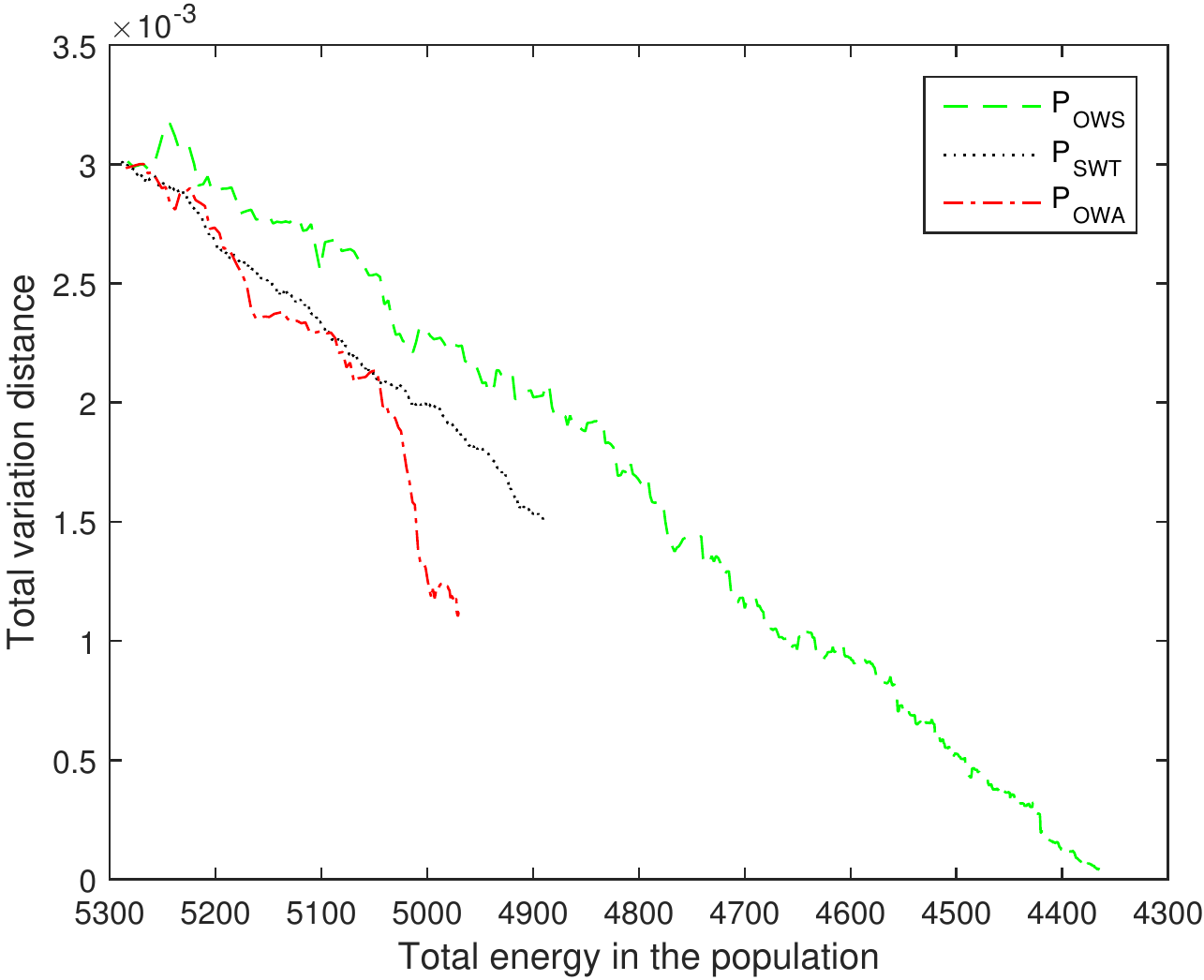}
                \caption{Efficiency, $\beta = 0.2$.}
                \label{fig:efficiency1}
        \end{subfigure}%

        \begin{subfigure}[b]{0.33\textwidth}
                \includegraphics[width=\textwidth]{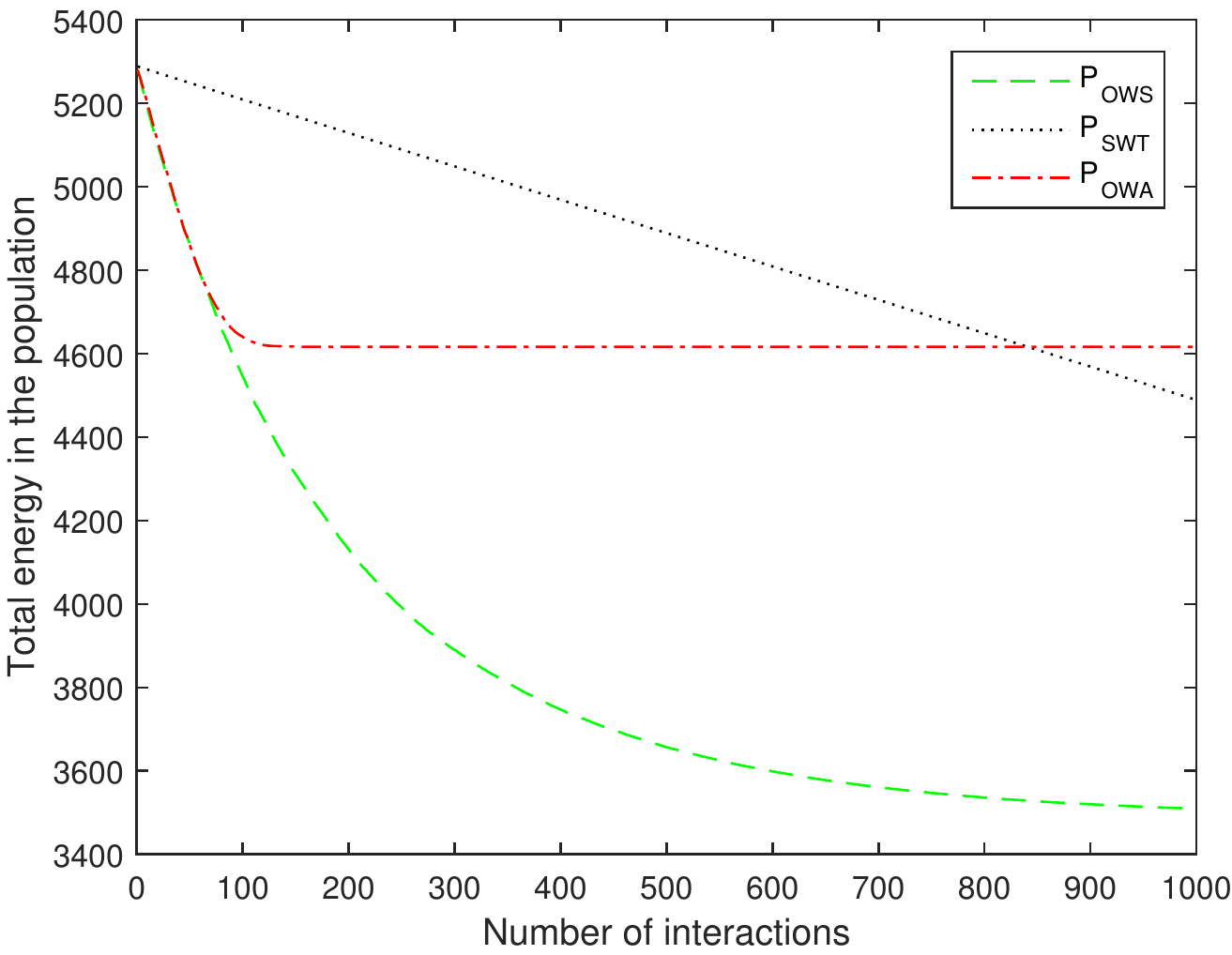}
                \caption{Energy loss, $\beta = 0.4$.}
                \label{fig:loss2}
        \end{subfigure}%
        \begin{subfigure}[b]{0.33\textwidth}
                \includegraphics[width=\textwidth]{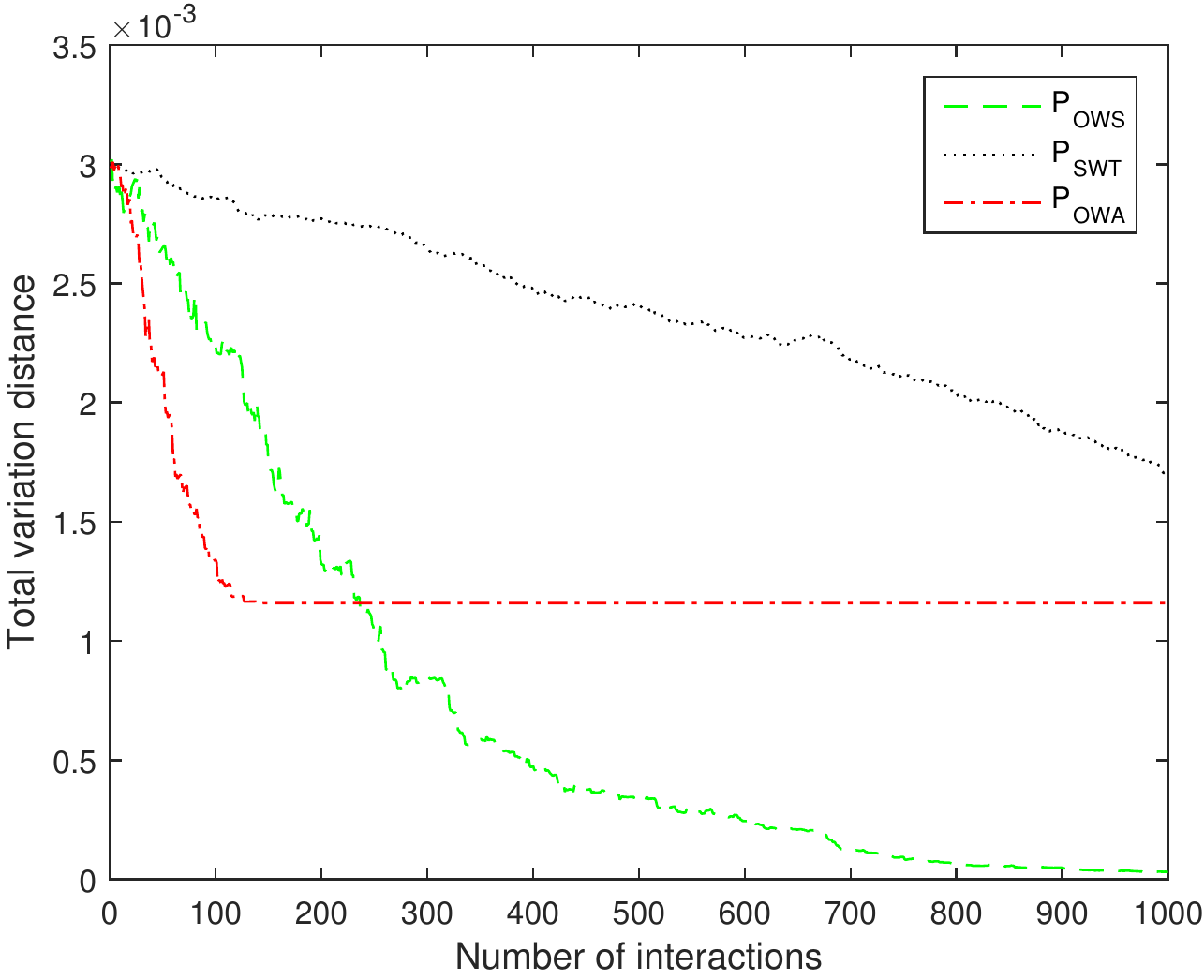}
                \caption{Weighted energy balance, $\beta = 0.4$.}
                \label{fig:balance2}
        \end{subfigure}%
        \begin{subfigure}[b]{0.33\textwidth}
                \includegraphics[width=\textwidth]{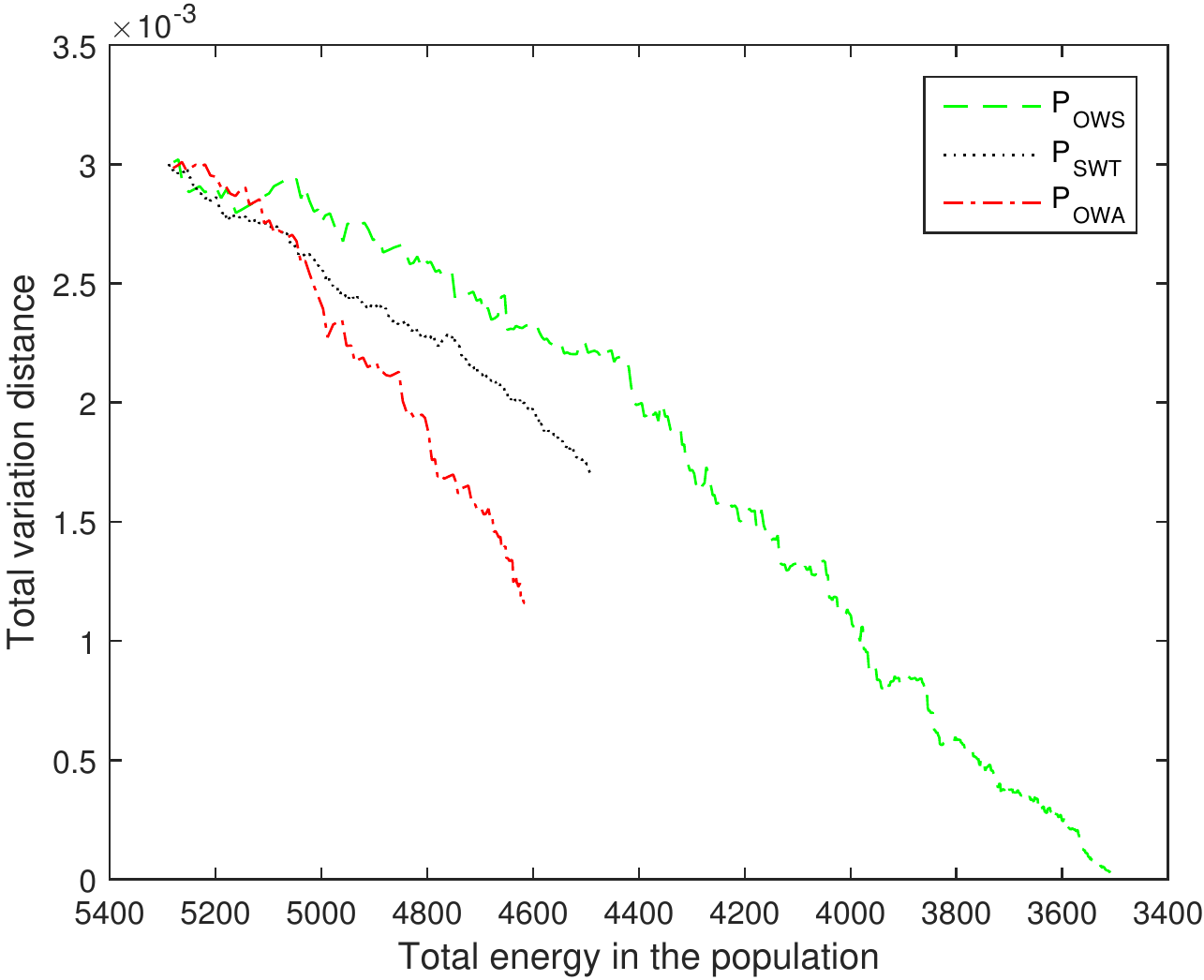}
                \caption{Efficiency, $\beta = 0.4$.}
                \label{fig:efficiency2}
        \end{subfigure}%
        
        \begin{subfigure}[b]{0.33\textwidth}
                \includegraphics[width=\textwidth]{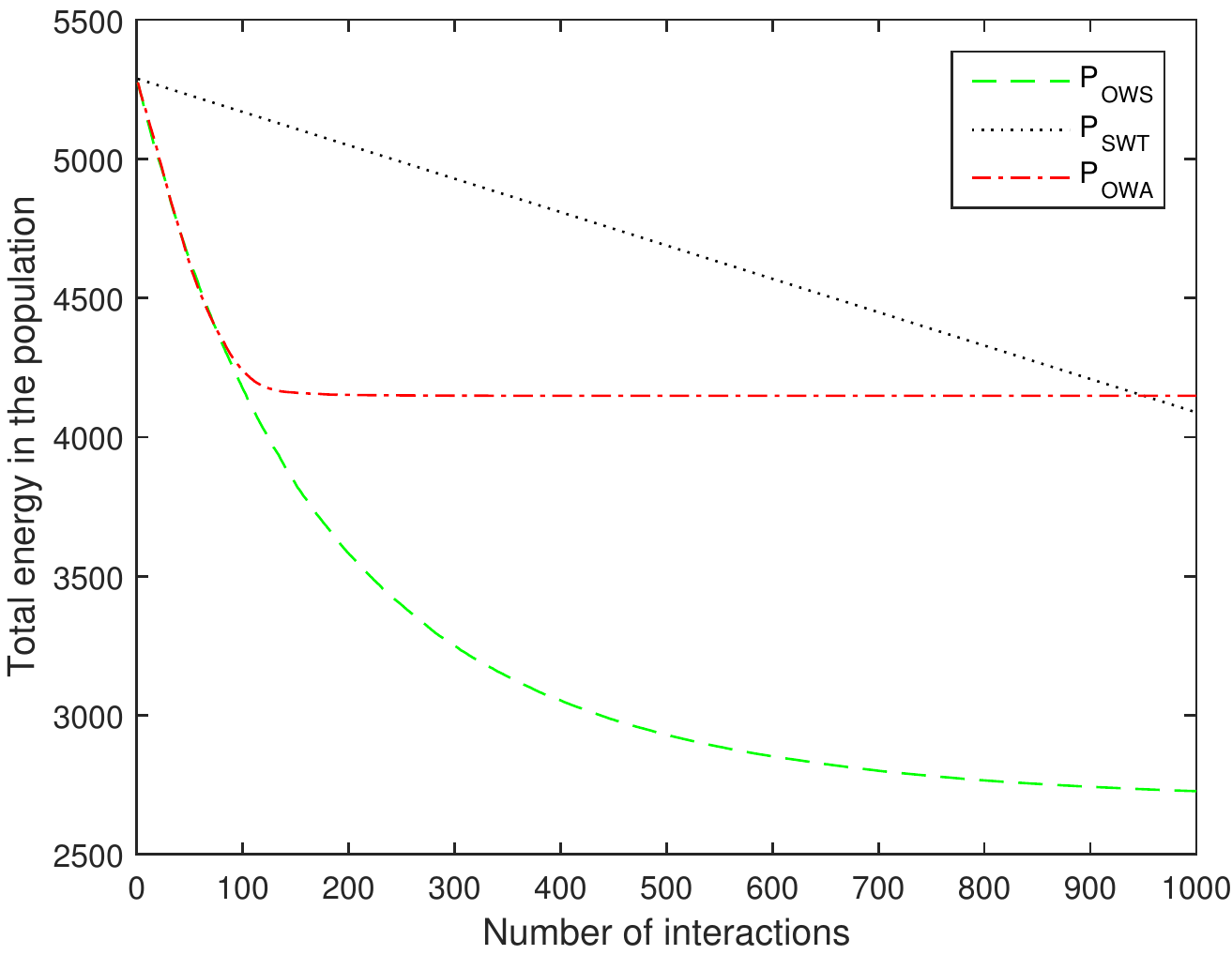}
                \caption{Energy loss, $\beta = 0.6$.}
                \label{fig:loss3}
        \end{subfigure}%
        \begin{subfigure}[b]{0.33\textwidth}
                \includegraphics[width=\textwidth]{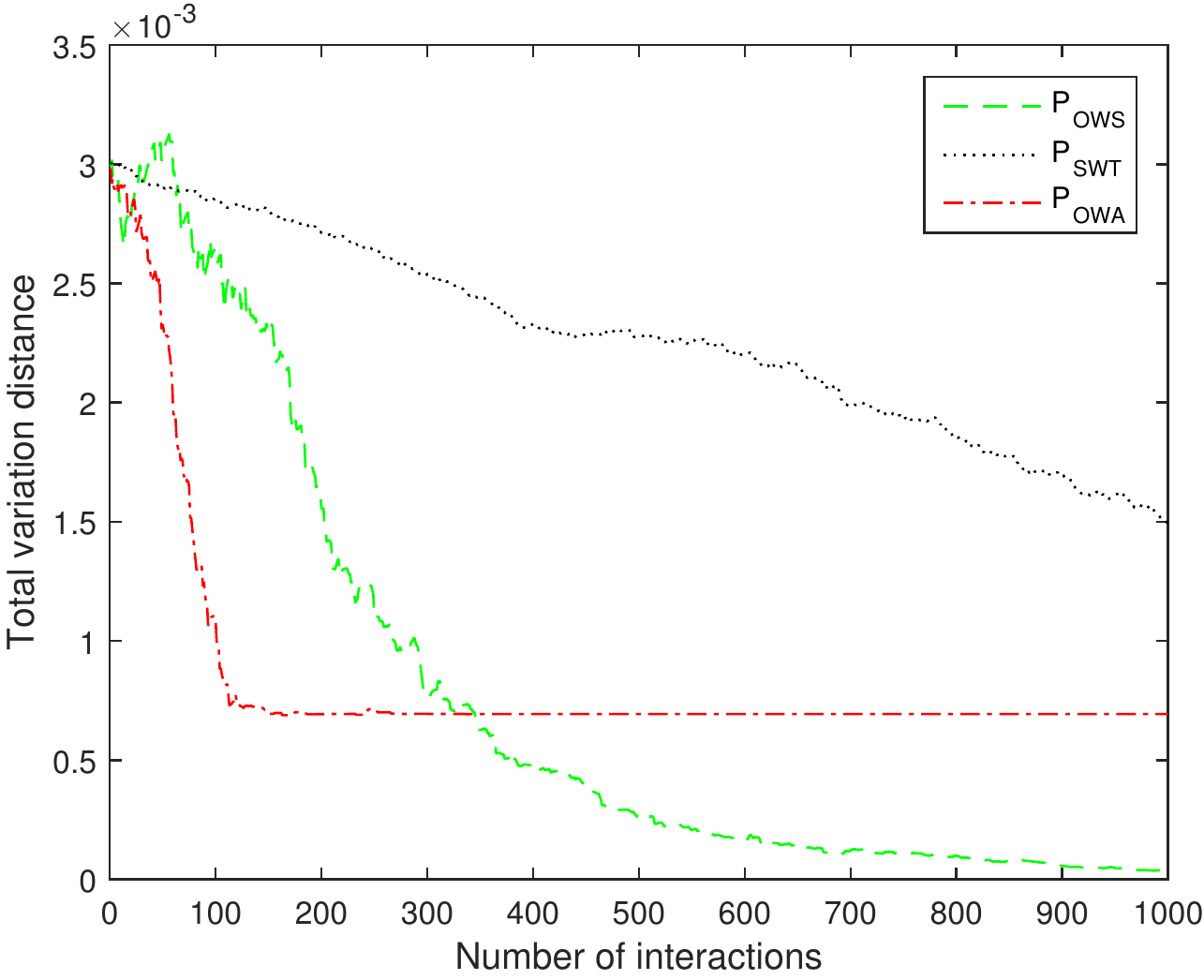}
                \caption{Weighted energy balance, $\beta = 0.6$.}
                \label{fig:balance3}
        \end{subfigure}%
        \begin{subfigure}[b]{0.33\textwidth}
                \includegraphics[width=\textwidth]{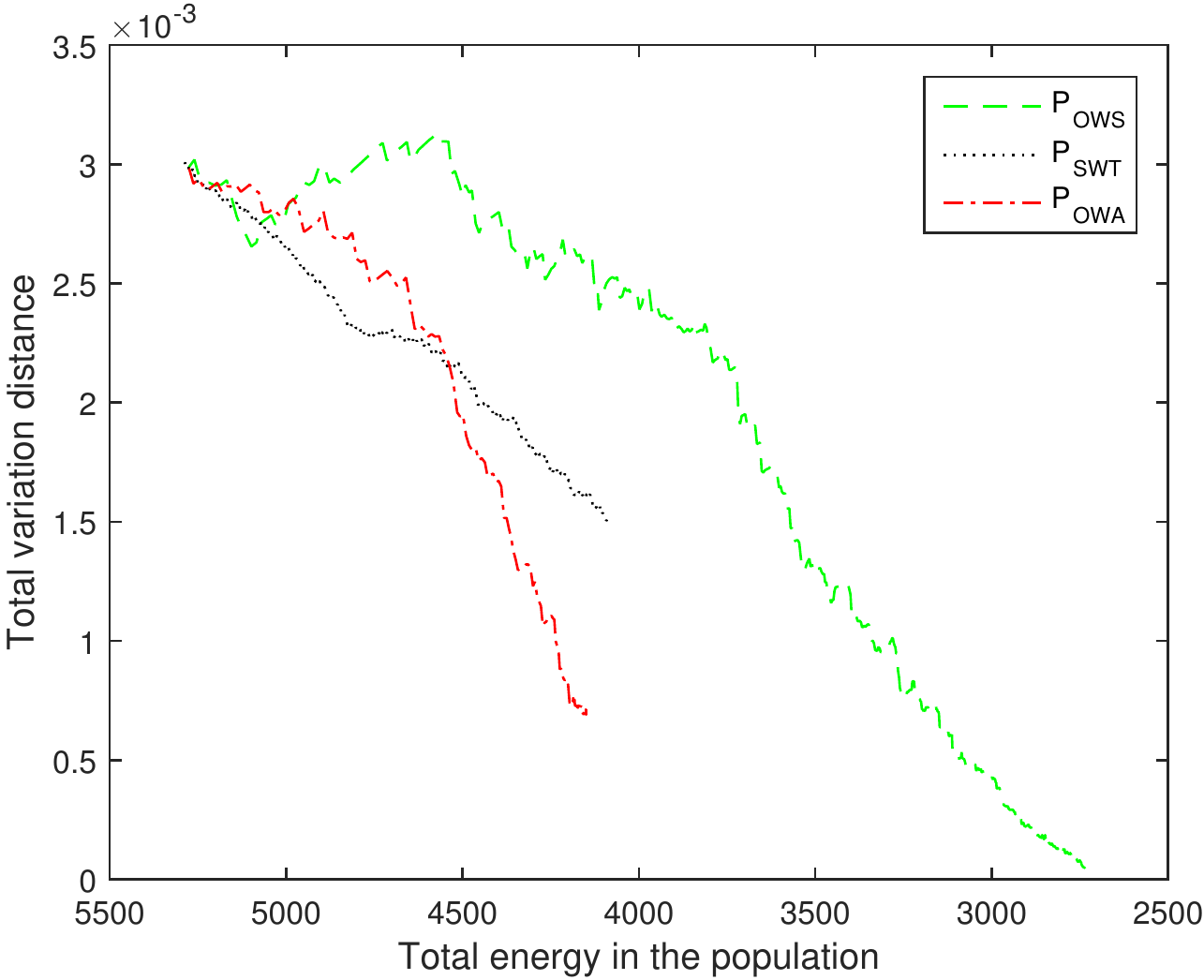}
                \caption{Efficiency, $\beta = 0.6$.}
                \label{fig:efficiency3}
        \end{subfigure}%
        
        \begin{subfigure}[b]{0.33\textwidth}
                \includegraphics[width=\textwidth]{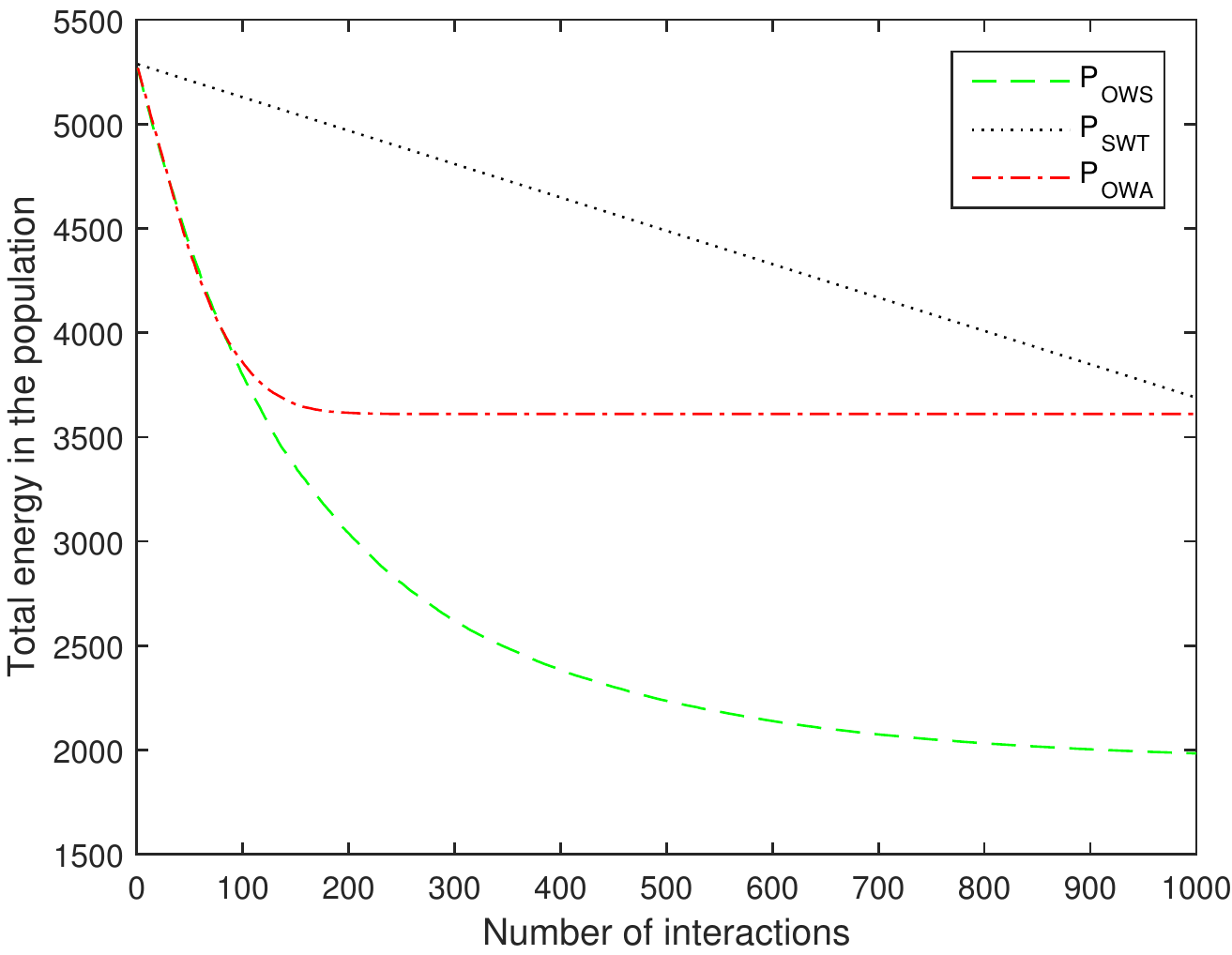}
                \caption{Energy loss, $\beta = 0.8$.}
                \label{fig:loss4}
        \end{subfigure}%
        \begin{subfigure}[b]{0.33\textwidth}
                \includegraphics[width=\textwidth]{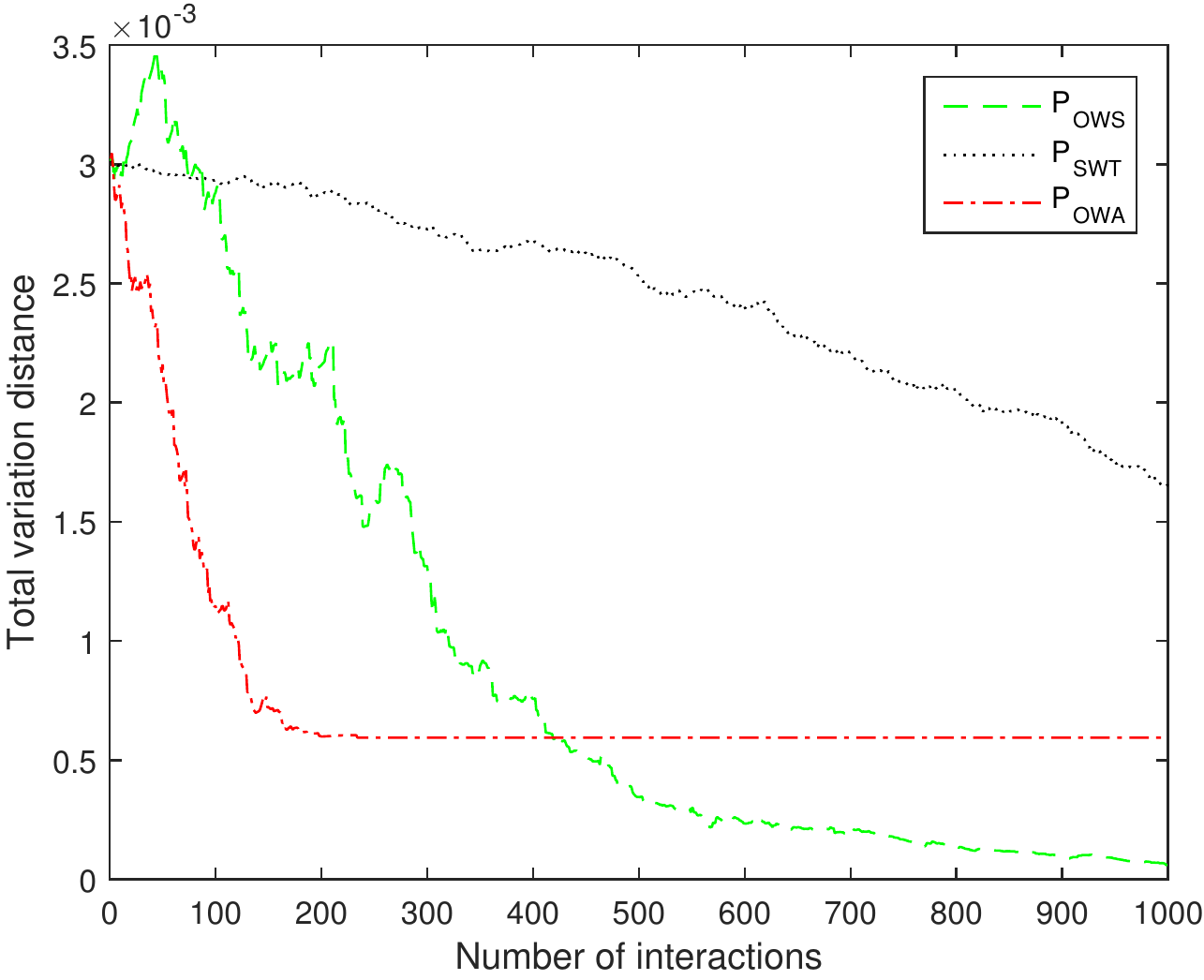}
                \caption{Weighted energy balance, $\beta = 0.8$.}
                \label{fig:balance4}
        \end{subfigure}%
        \begin{subfigure}[b]{0.33\textwidth}
                \includegraphics[width=\textwidth]{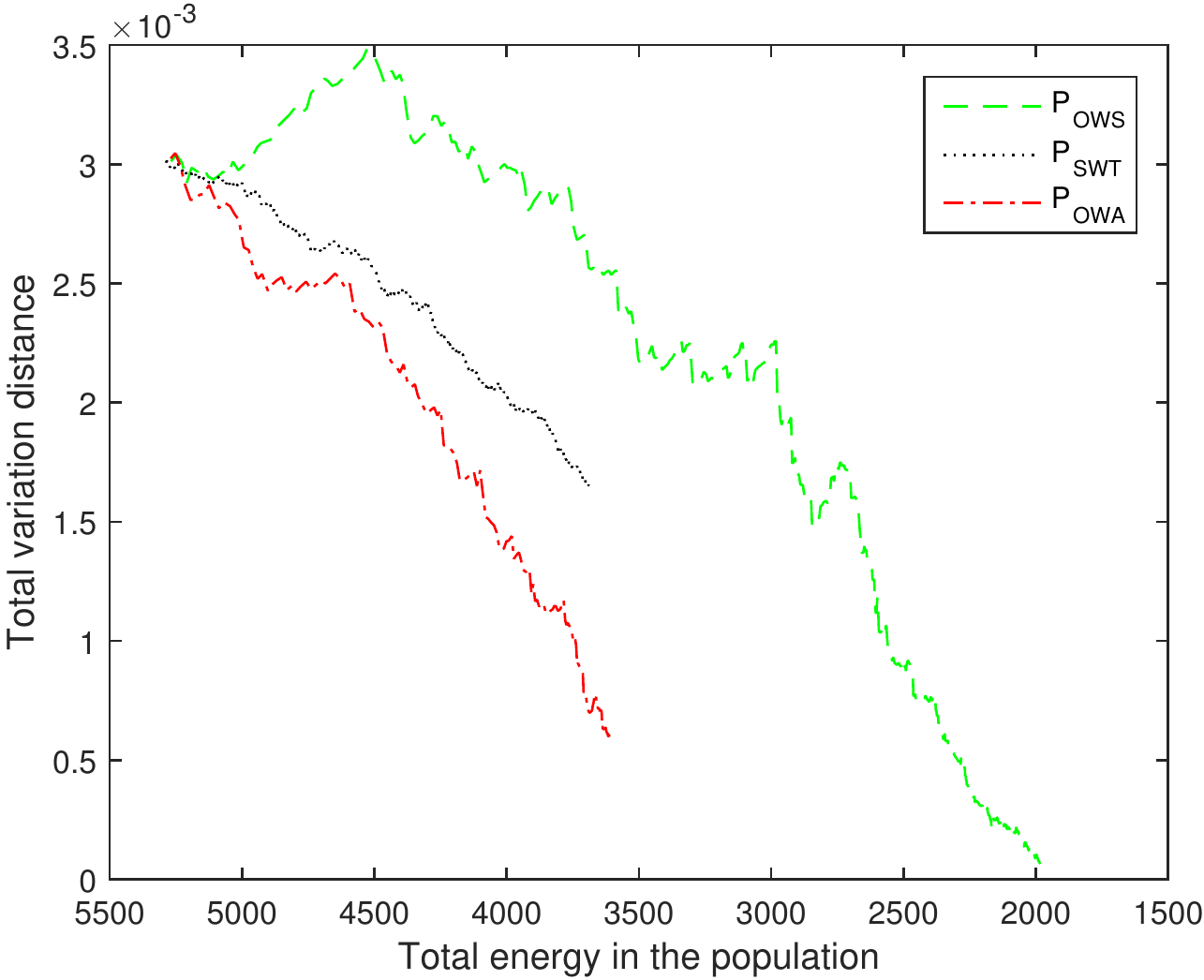}
                \caption{Efficiency, $\beta = 0.8$.}
                \label{fig:efficiency4}
        \end{subfigure}%
        \caption{Comparison of the three protocols for various metrics and different values of $\beta$.}
        \label{fig:comparison}
\end{figure*}

We conducted simulations in order to evaluate our methods, using Matlab R2014b. We compared the protocols ${\cal P}_{\text{OWS}}$, ${\cal P}_{\text{SWT}}$ and ${\cal P}_{\text{OWA}}$ by conducting experiment runs of $1.000$ useful interactions, where the nodes to interact are selected by a probabilistic scheduler. We assign an initial energy level value to every agent of a population consisting of $|m| = 100$ agents uniformly at random, with minimum and maximum battery cell capacityof $1$ and $100$ units of energy correspondingly (uniformly distributed at random). We create a highly unbalanced instance of weights distribution by splitting the weights in two categories, very high weights (critical agents) and very low weights (non-critical agents). The constant $\beta$ of the loss function is set to four different values, as different energy losses might lead to different performance (see Fig.~\ref{fig:beta}). For statistical smoothness, we repeat each experiment $100$ times. The statistical analysis of the findings (the median, lower and upper quartiles, outliers of the samples) demonstrate very high concentration around the mean, so in the following figures we only depict average values.

\subsection{The impact of $\beta$}

Different loss functions $L(\varepsilon)$ lead to different performance of the interaction protocols, both when running the same protocol and when comparing different protocols. Regarding the impact of different values of the $\beta$ constant on the same protocol, an example is shown in Fig.~\ref{fig:beta}. The total variation distance w.r.t.~the remaining energy in the population is shown. We ran the ${\cal P}_{\text{SWT}}$ protocol for values $0.2, 0.4$, $0.6$ and $0.8$. The results clearly show that the bigger the $\beta$, the larger the total variation distance for a given total level of energy in the population. For this reason, we decided to comparatively evaluate our protocols for different values of $\beta$, as shown in Fig.~\ref{fig:comparison}. As for the impact on different protocols, if we observe Figs.~\ref{fig:loss1}, \ref{fig:loss2}, \ref{fig:loss3} and \ref{fig:loss4} carefully, we can see that, for the same total initial energy and number of useful interactions, when the $\beta$ constant and consequently the energy loss increases, the rate of total energy loss increases as well.

\subsection{Energy loss}

Figs.~\ref{fig:loss1}, \ref{fig:loss2}, \ref{fig:loss3} and \ref{fig:loss4} are depicting the total energy of the population over time, for $1.000$ useful agent interactions of the three protocols ${\cal P}_{\text{OWS}}$, ${\cal P}_{\text{SWT}}$ and ${\cal P}_{\text{OWA}}$. Each protocol's behavior is similar, regardless of the value of $\beta$, but with higher losses when $\beta$ increases. The energy loss rate for ${\cal P}_{\text{OWS}}$ and ${\cal P}_{\text{OWA}}$ is high in the beginning, until a point of time after which it significantly drops. This is explained by the fact that, in the beginning, both protocols perform interactions of high energy transfer amounts $\varepsilon$ which lead to high $L(\varepsilon)$. After this initial phase, interactions with large energy transfers are very rare, and so ${\cal P}_{\text{OWS}}$ and ${\cal P}_{\text{OWA}}$ perform energy transfers of very small $\varepsilon$, forcing the energy loss rate to drop. ${\cal P}_{\text{SWT}}$ has a smoother, linear energy loss rate, since $\varepsilon$ is a very small fixed value. 

\subsection{Weighted energy balance}

Useful conclusions about weighted energy balance of the population can be derived from Figures.~\ref{fig:balance1}, \ref{fig:balance2}, \ref{fig:balance3} and \ref{fig:balance4}, where we can see how total variation distance changes over time. A first remark is that the protocols are balancing the available energy in the population in an analogous rate to the energy loss rate. Since a better weighted energy balance is expressed by lower values of total variation distance, it is apparent that eventually the best balance after $1.000$ useful interactions is provided by ${\cal P}_{\text{OWS}}$. However, note that this is a conclusion regarding only the weighted energy balance, not taking into account the losses from the charging procedure. ${\cal P}_{\text{OWS}}$ eventually achieves weighted energy balance fast, because much energy is being gradually lost by the network and the protocol is managing significantly smaller amounts of energy. Better balance does not necessarily lead to higher overall efficiency, w.r.t.~energy loss. If we take a better look at the weighted energy balance figures, we observe that even if the total variation distance follows a decreasing pattern, it is not strictly decreasing. This is natural, since many interactions can temporarily lead to a worse weighted energy balance in the population due to sharp changes in the distribution of total energy (see also the discussion in Section~\ref{sec:lossy}).

\subsection{Convergence time}

The time that each protocol needs for balancing the available energy in the population, is not a negligible factor. Quick balancing leads to transfers of significantly smaller amounts of energy among agents and consequently to lower energy losses. On the other hand, in order to achieve quick balancing, in some cases there has been already much energy loss due to frequent lossy interactions. In Fig.~\ref{fig:comparison}, we can see that ${\cal P}_{\text{OWA}}$ is the fastest to achieve a stable level of weighted energy balance in the population, as opposed to ${\cal P}_{\text{SWT}}$, which is wasteful in terms of running time. ${\cal P}_{\text{OWS}}$ performance with respect to time lies somewhere in between the two other protocols, since it is able to conduct all types of interactions (unlike ${\cal P}_{\text{OWA}}$ in which only some interactions are allowed and ${\cal P}_{\text{SWT}}$ which performs only interactions of small $\varepsilon$).

\subsection{Overall efficiency}

We measure the overall efficiency of a protocol by taking into account both energy losses and weighted energy balance in the population. This combination of the two crucial properties is shown in Figs.~\ref{fig:efficiency1}, \ref{fig:efficiency2}, \ref{fig:efficiency3} and \ref{fig:efficiency4}, where ${\cal P}_{\text{SWT}}$ and ${\cal P}_{\text{OWA}}$ clearly outperform ${\cal P}_{\text{OWS}}$, most of the time. More specifically, although ${\cal P}_{\text{OWS}}$ achieves very good balance quickly, the impact of energy loss affect very negatively it's overall performance. This is due to the fact that for the same amount of total energy in the population, ${\cal P}_{\text{SWT}}$ and ${\cal P}_{\text{OWA}}$ achieve better total variation distance than ${\cal P}_{\text{OWS}}$. It is also clear than eventually, ${\cal P}_{\text{OWA}}$ outperforms both ${\cal P}_{\text{OWS}}$ and ${\cal P}_{\text{SWT}}$ when agent interactions are planned by the probabilistic scheduler. Furthermore, it is much faster than ${\cal P}_{\text{SWT}}$ in terms of the number of useful interactions.

\section{Conclusion}
In this paper we apply interactive wireless charging in populations of resource-limited, mobile agents that abstract distributed portable devices. We provide a model in which the agents are capable of achieving peer-to-peer wireless energy transfer and can act both as energy transmitters and as energy harvesters. We consider the cases of loss-less and lossy wireless energy transfer and, in the former case, we prove a tight upper bound on the time that is needed to reach weighted energy balance in the population. Finally, we design and evaluate three interaction protocols which opt for creating weighted energy balance among the agents of the population.


\balance
\bibliographystyle{elsarticle-num}

\end{document}